\documentclass[fullpage, 12pt, a4paper, doublespacing]{article}

\usepackage[utf8]{inputenc}
\usepackage{pgfplots}
\usepackage{pgfplotstable}
\usepackage{verbatim}
\usepackage{booktabs}
\usepackage{pdfpages}
\usepackage{url}

\usepackage{graphicx}
\usepackage{amsmath}
\usepackage{subfigure}
\usepackage{natbib}
\usepackage{bm}

\definecolor{transfertoserver}{HTML}{D7191C}
\definecolor{database}{HTML}{FDAE61}
\definecolor{transfertoclient}{HTML}{ABDDA4}
\definecolor{rendering}{HTML}{2B83BA}

\usepackage{psfrag}

\newcommand{\varphibold}{\mbox{\boldmath$\varphi$}}
\newcommand{\xibold}{\mbox{\boldmath$\xi$}}
\newcommand{\be}{\begin{equation}}
\newcommand{\ee}{\end{equation}}

\def\bkRrm{{\rm I\kern-.17em R}}

\title{Dance Hit Song Prediction}
 \author{Dorien Herremans$^{a}$\thanks{Corresponding author. Email: dorien.herremans@uantwerpen.be
 \vspace{6pt}}, David Martens$^{b}$ and Kenneth Sörensen$^{a}$\vspace{6pt}\\ \emph{$^{a}$ANT/OR, University of Antwerp Operations Research Group}\\ \emph{$^{b}$Applied Data Mining Research Group, University of Antwerp}\\ \emph{Prinsstraat 13, B-2000 Antwerp}}

\DeclareOldFontCommand{\bf}{\normalfont\bfseries}{\mathbf}
\usepackage{fancyhdr}

\begin{document}
	
	\pagestyle{fancy}
	\chead{ }
	\fancyhf{} % sets both header and footer to nothing
	\renewcommand{\headrulewidth}{0pt}
	\cfoot{\emph{Preprint accepted for publication in: Herremans D., Martens D, Sörensen K.. 2014. Journal of New Music research. Vol. 43 - Special Issue on Music and Machine Learning, pp291-302.} \\ \vspace{0.4cm}\thepage }

\maketitle

\begin{abstract}
Record companies invest billions of dollars in new talent around the globe each year. Gaining insight into what actually makes a \emph{hit} song would provide tremendous benefits for the music industry. In this research we tackle this question by focussing on the \emph{dance} hit song classification problem. A database of dance hit songs from 1985 until 2013 is built, including basic musical features, as well as more advanced features that capture a temporal aspect. A number of different classifiers are used to build and test dance hit prediction models. The resulting best model has a good performance when predicting whether a song is a ``top 10'' dance hit versus a lower listed position. 
\end{abstract}

\section{Introduction}

In 2011 record companies invested a total of 4.5 billion in new talent worldwide~\citep{ifpi2012}. Gaining insight into what actually makes a song a hit would provide tremendous benefits for the music industry. This idea is the main drive behind the new research field referred to as ``Hit song science'' which ~\cite{pachet2012hit} define as ``an emerging field of science that aims at predicting the success of songs before they are released on the market''.

There is a large amount of literature available on song writing techniques~\citep{braheny2007, Webb1999}. Some authors even claim to teach the reader how to write \emph{hit} songs~\citep{molly2008, jack2000}. Yet very little research has been done on the task of automatic prediction of hit songs or detection of their characteristics. 

The increase in the amount of digital music available online combined with the evolution of technology has changed the way in which we listen to music. In order to react to new expectations of listeners who want searchable music collections, automatic playlist suggestions, music recognition systems etc., it is essential to be able to retrieve information from music~\citep{casey2008content}. This has given rise to the field of Music Information Retrieval (MIR), a multidisciplinary domain concerned with retrieving and analysing multifaceted information from large music databases~\citep{downie2003music}. 

Many MIR systems have been developed in recent years and applied to a range of different topics such as automatic classification per genre~\citep{tzanetakis2002musical}, cultural origin~\citep{whitman2002combining}, mood~\citep{laurier2008multimodal}, composer~\citep{herremans2013}, instrument~\citep{essid2006musical}, similarity~\citep{schnitzer2009filter}, etc. An extensive overview is given by~\cite{fu2011survey}. Yet, as it appears, the use of MIR systems for hit prediction remains relatively unexplored. 

The first exploration into the domain of hit science is due to~\cite{dhanaraj2005automatic}. They used acoustic and lyric-based features to build support vector machines (SVM) and boosting classifiers to distinguish top 1 hits from other songs in various styles. Although acoustic and lyric data was only available for 91 songs, their results seem promising. The study does however not provide details about data gathering, features, applied methods and tuning procedures. 

Based on the claim of the unpredictability of cultural markets made by~\cite{salganik2006experimental},~\cite{pachet2008hit} examined the validity of this claim on the music market. Based on a dataset they were not able to develop an accurate classification model for low, medium or high popularity based on acoustic and human features. They suggest that the acoustic features they used are not informative enough to be used for aesthetic judgements and suspect that the previously mentioned study~\citep{dhanaraj2005automatic} is based on spurious data or biased experiments. %how hit determined in their experiment?

~\cite{borg83} draw similar conclusions as~\cite{pachet2008hit}. They tried to predict the popularity of music videos based on their YouTube view count by training support vector machines but were not successful. 

Another experiment was set up by~\cite{ni2011hit}, who claim to have proven that hit song science is once again a science. They were able to obtain more optimistic results by predicting if a song would reach a top 5 position on the UK top 40 singles chart compared to a top 30-40 position. The shifting perceptron model that they built was based on thus far novel audio features mostly extracted from The Echo Nest\footnote{\url{echonest.com}}. Though they describe the features they used on their website~\citep{jehan2012}, the paper is very short and does not disclose a lot of details about the research such as data gathering, preprocessing, detailed description of the technique used or its implementation.

In this research accurate models are built to predict if a song is a top 10 dance hit or not. For this purpose, a dataset of dance hits including some unique audio features is compiled. Based on this data different efficient models are built and compared. To the authors' knowledge, no previous research has been done on the dance hit prediction problem.

In the next section, the dataset used in this paper is elaborately discussed. In Section~\ref{sec:vis} the data is visualized in order to detect some temporal patterns. Finally, the experimental setup is described and a number of models are built and tested.

\section{Dataset}

The dataset used in this research was gathered in a few stages. The first stage involved determining which songs can be considered as hit songs versus which songs cannot. Secondly, detailed information about musical features was obtained for both aforementioned categories.

\subsection{Hit Listings}

Two hit archives available online were used to create a database of dance hits (see Table~\ref{tab:hits}). The first one is the singles dance archive from the Official Charts Company (OCC)\footnote{\url{officialcharts.com}}. The Official Charts Company is operated by both the British Phonographic Industry and the Entertainment Retailers Association ERA. Their charts are produced based on sales data from retailers through market researcher Millward Brown. 
The second source is the singles dance archive from Billboard (BB)\footnote{\url{billboard.com}}. Billboard is one of the oldest magazines in the world devoted to music and the music industry.

The information was parsed from both websites using the Open source Java html parser library JSoup~\citep{houstoninstant2013} and resulted in a dataset of 21,692 (7,159 + 14,533) listings with 4 features: song title, artist, position and date. A very small number of hit listings could not be parsed and these were left out of the dataset. The peak chart position for each song was computed and added to the dataset as a fifth feature. Table~\ref{tab:example} shows an example of the dataset at this point.

% previously called the Chart Information Network (CIN) and then The Official UK Charts Company, compiles various "official" UK record charts, including the UK Singles Chart, the UK Albums Chart, and the UK Official Download Chart, as well as genre-specific and music video charts.
% The OCC produces its charts by gathering and combining sales data from retailers through market researchers Millward Brown. OCC claims to cover 99\% of the singles market and 95\% of the album market, and aims to collect data from any retailer who sells more than 100 chart items per week.[1]

% 
% Since 1 July 1997, CIN and then OCC have compiled the official charts. Prior to this date, the charts were produced by a succession of market research companies, beginning with the British Market Research Bureau in 1969, and later by Gallup. Before the production of the "official" charts, various less comprehensive charts were produced, most notably by the NME, which began its chart in 1952; some of these older charts (including NME's earliest singles charts) are now part of the official OCC canon.

 \begin{table}
 \centering 
 \caption{Hit listings overview.}
 \begin{tabular}{l|cc}
  \toprule
   & OCC & BB\\
   \midrule
Top & 40 & 10\\
Date range & 10/2009--3/2013 & 1/1985--3/2013\\
Hit listings & 7,159 & 14,533\\
Unique songs & 759 & 3,361\\
\bottomrule
  \end{tabular}
\label{tab:hits}
\end{table}

\begin{table}
 \centering
 \caption{Example of hit listings before adding musical features.}
 \small
 \begin{tabular}{llccc}
   \toprule
Song title & Artist & Position & Date & Peak position\\
\midrule
Harlem Shake 	& Bauer & 2	& 09/03/13 &		1	\\
Are You Ready For Love	&Elton John&40	&08/12/12	&	34\\
The Game Has Changed	&Daft Punk	&	32&18/12/10&	32\\
\dots&&&&\\
\bottomrule
 \end{tabular}
 \label{tab:example}
\end{table}

\subsection{Feature Extraction And Calculation}

The Echo Nest\footnote{\url{echonest.com}} was used in order to obtain musical characteristics for the song titles obtained in previous subsection. The Echo Nest is the world's leading music intelligence company and has over a trillion data points on over 34 million songs in its database. Its services are used by industry leaders such as Spotify, Nokia, Twitter, MTV, EMI and more~\citep{echo2013}.~\cite{BertinMahieux2011} used The Echo Nest to build The One Million Song dataset, a very large freely available dataset that offers a collection of audio features and meta-information for a million contemporary popular songs. 

In this research The Echo Nest was used to build a new database mapped to the hit listings. The Open Source java client library jEN for the Echo Nest developer API was used to query the songs~\citep{jen2013}. Based on the song title and artist name, The Echo Nest database and Analyzer were queried for each of the parsed hit songs. After some manual and java-based corrections for spelling irregularities (e.g., Featuring, Feat, Ft.) data was retrieved for 697 out of 759 unique songs from the OCC hit listings and 2,755 out of 3,361 unique songs from the BB hit listings. The %668 
songs with missing data were removed from the dataset. The extracted features can be divided into three categories: meta-information, basic features from The Echo Nest Analyzer and temporal features. %The resulting database will be made available online. 

% \includegraphics[scale=0.5]{jen.png}
% OCC
%  7159 hits (759 unique songs)
% 7090 with data (697 unique songs)
% % \includegraphics[scale=0.5]{occ.png}
%  BB
%  14533 hits (3361 unique songs)
%  12711 with data (2755 unique songs)
% % \includegraphics[scale=0.2]{bb.jpg}

\subsubsection{Meta-Information}

 The first category is \emph{meta-information} such as artist location, artist familiarity, artist hotness, song hotness etc. This is descriptive information about the song, often not related to the audio signal itself. One could follow the statement of IBM's Bob Mercer in 1985 ``There is no data like more data''~\citep{jelinek2005some}. Yet, for this research, the meta-information is discarded when building the classification models. In this way, the model can work with unknown songs, based purely on audio signals. 
 
 \subsubsection{Basic Analyzer Features}
 The next category consists of \emph{basic features} extracted by The Echo Nest Analyzer~\citep{jehan2012}. Most of these features are self-explanatory, except for \emph{energy} and \emph{danceability}, of which The Echo Nest did not yet release the formula.

\begin{description}
 \item[Duration] Length of the track in seconds.
 \item[Tempo] The average tempo expressed in beats per minute (bpm).
 \item[Time signature] A symbolic representation of how many beats there are in each bar. 
 \item[Mode] Describes if a song's modality is major (1) or minor (0).
 \item[Key] The estimated key of the track, represented as an integer.
 \item[Loudness] The loudness of a track in decibels (dB), which correlates to the psychological perception of strength (amplitude).
 \item[Danceability] Calculated by The Echo Nest, based on beat strength, tempo stability, overall tempo, and more. %cite blog? running with data?
 \item[Energy] Calculated by The Echo Nest, based on loudness and segment durations.
\end{description}

A more detailed description of these Echo Nest features is given by~\cite{jehan2012}.

\subsubsection{Temporal Features}
A third category of features was added to incorporate the \emph{temporal aspect} of the following basic features offered by the Analyzer:

\begin{description}
 \item[Timbre] A 12-dimensional vector which captures the tone colour for each segment of a song. A segment is a sound entity (typically under a second) relatively uniform in timbre and harmony.
 \item[Beatdiff] The time difference between subsequent beats. 
\end{description}

Timbre is a very perceptual feature that is sometimes referred to as tone colour. In The Echo Nest, 13 basis vectors are available that are derived from the principal components analysis (PCA) of the auditory spectrogram~\citep{jehan2005creating}. The first vector of the PCA is referred to as loudness, as it is related to the amplitude. The following 12 basis vectors are referred to as the timbre vectors. The first one can be interpreted as brightness, as it emphasizes the ratio of high frequencies versus low frequencies, a measure typically correlated to the ``perceptual'' quality of brightness. The second timbre vector has to do with flatness and narrowness of sound (attenuation of lowest and highest frequencies). The next vector represents the emphasis of the attack (sharpness)~\citep{echo2013}. The timbre vectors after that are harder to label, but can be understood by the spectral diagrams given by~\cite{jehan2005creating}. 

In order to capture the temporal aspect of timbre throughout a song~\cite{schindler2012capturing} introduce a set of derived features. They show that genre classification can be significantly improved by incorporating the statistical moments of the 12 segment timbre descriptors offered by The Echo Nest. In this research the statistical moments were calculated together with some extra descriptive statistics: mean, variance, skewness, kurtosis, standard deviation, 80th percentile, min, max, range and median.

~\cite{score2013} introduce a variable called Beat CV in their model, which refers to the variation of the time between the beats in a song. In this research, the temporal aspect of the time between beats (beatdiff) is taken into account in a more complete way, using all the descriptive statistics from the previous paragraph.

After discarding the meta-information, the resulting dataset contained 139 usable features. In the next section, these features were analysed to discover their evolution over time.

% Skewness (3rd moment) (lean of the distribution)
% Kurtosis (4th moment) (peakedness of the distribution)
%  \includegraphics[scale=0.3]{echonest.jpg}

\section{Evolution Over Time}
\label{sec:vis}

The dominant music that people listen to in a certain culture changes over time. It is no surprise that a hit song from the 60s will not necessarily fit in the contemporary charts. Even if we limit ourselves to one particular style of hit songs, namely dance music, a strong evolution can be distinguished between popular 90s dance songs and this week's hit. In order to verify this statement and gain insight into how characteristics of dance music have changed, the Billboard dataset (BB) with top 10 dance hits from 1985 until now was analysed.

\begin{figure}[h]
\centering
  \includegraphics[scale=0.65]{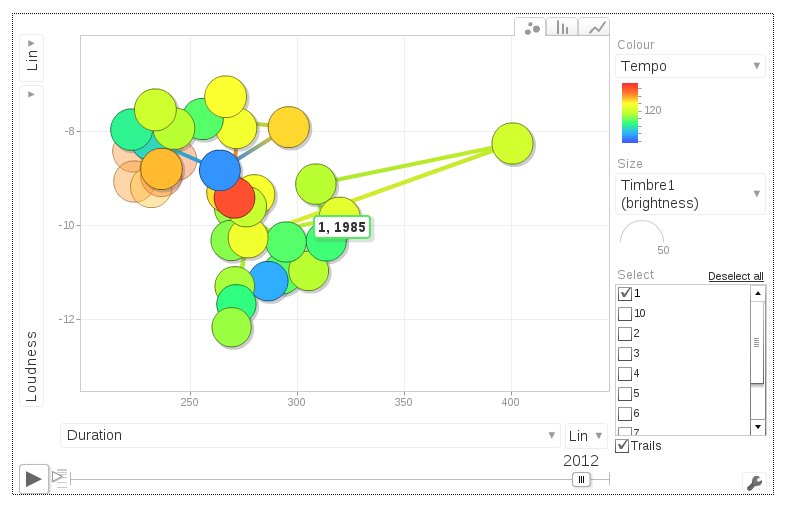}
  \caption{Motion chart visualising evolution of dance hits from 1985 until 2013$^5$.}
  \label{fig:motion}
\end{figure}

A dynamic chart was used to represent the evolution of four features over time~\citep{motion2013}. Figure~\ref{fig:motion} shows a screenshot of the Google motion chart\footnote{Interactive motion chart available at \url{http://antor.ua.ac.be/dance}} that was used to visualize the time series data. This graph integrates data mining and information
visualization in one discovery tool as it reveals interesting patterns and allows the user to control the visual presentation, thus following the recommendation made by~\cite{shneiderman2002inventing}. The x-axis shows the duration and the y-axis is the average loudness per year in Figure~\ref{fig:motion}. Additional dimensions are represented by the size of the bubbles (brightness) and the colour of the bubbles (tempo).

\begin{figure}[h]
\centering
\centerline{\subfigure[Duration]{   
  \includegraphics[scale=0.5]{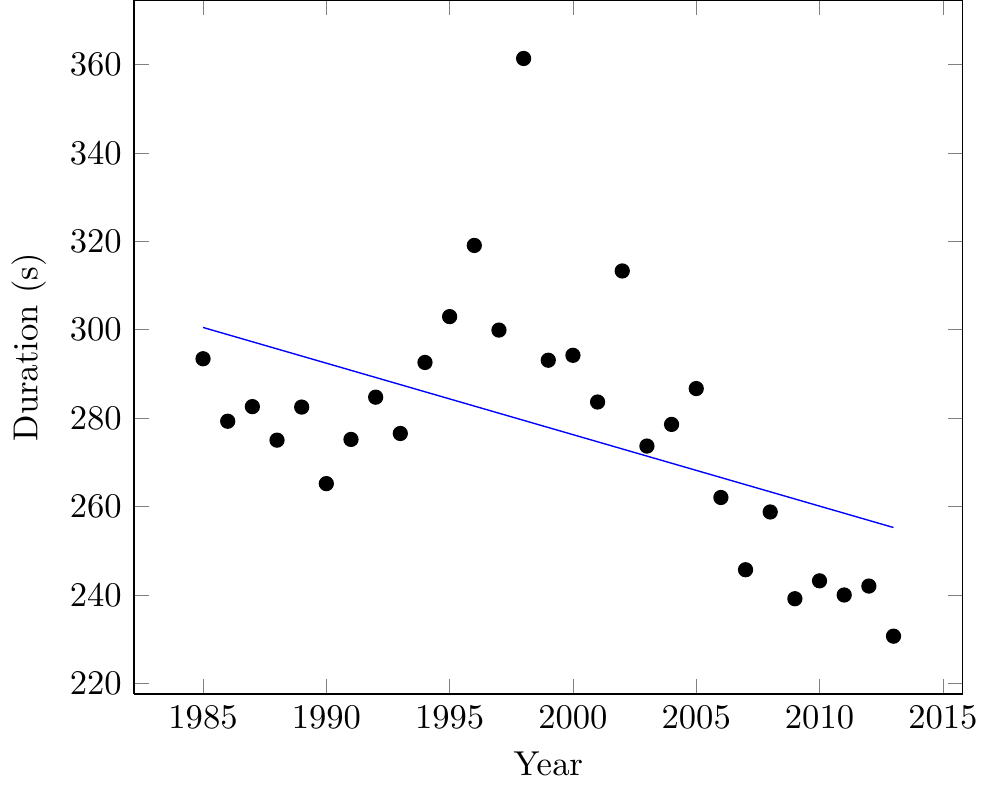}
}\subfigure[Tempo]{   
 \includegraphics[scale=0.5]{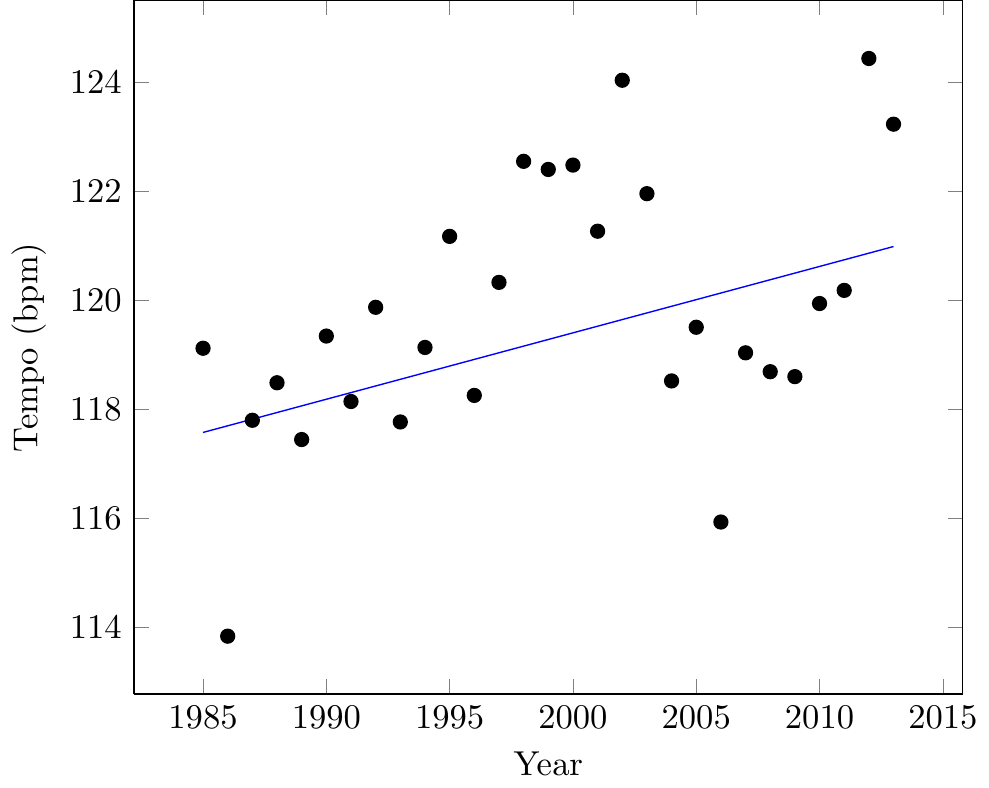}
}\subfigure[Loudness]{  
   \includegraphics[scale=0.5]{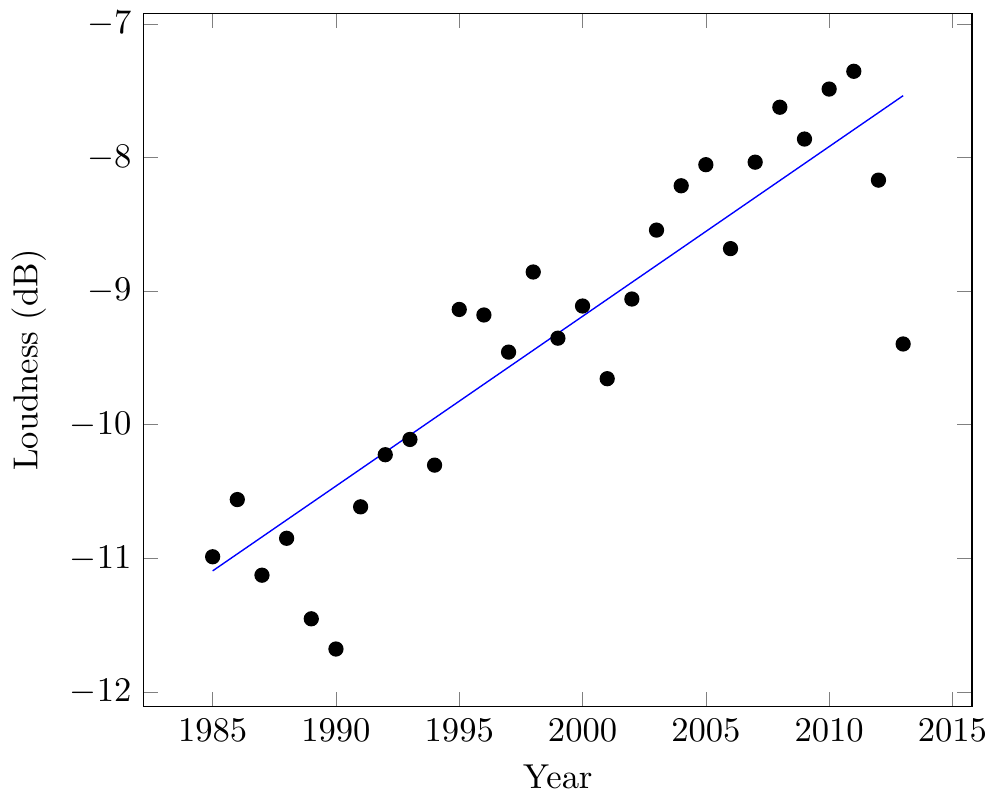}
}}
\centerline{\subfigure[Timbre 1 (mean)]{   
    \includegraphics[scale=0.5]{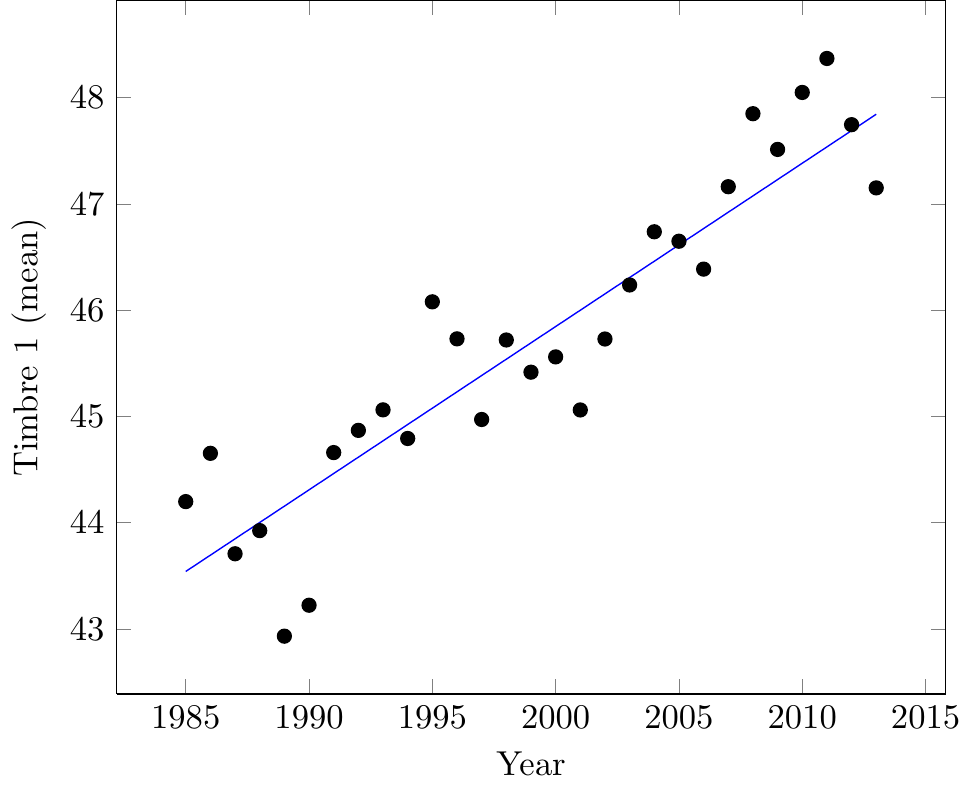} } 
\subfigure[Energy]{
 \includegraphics[scale=0.5]{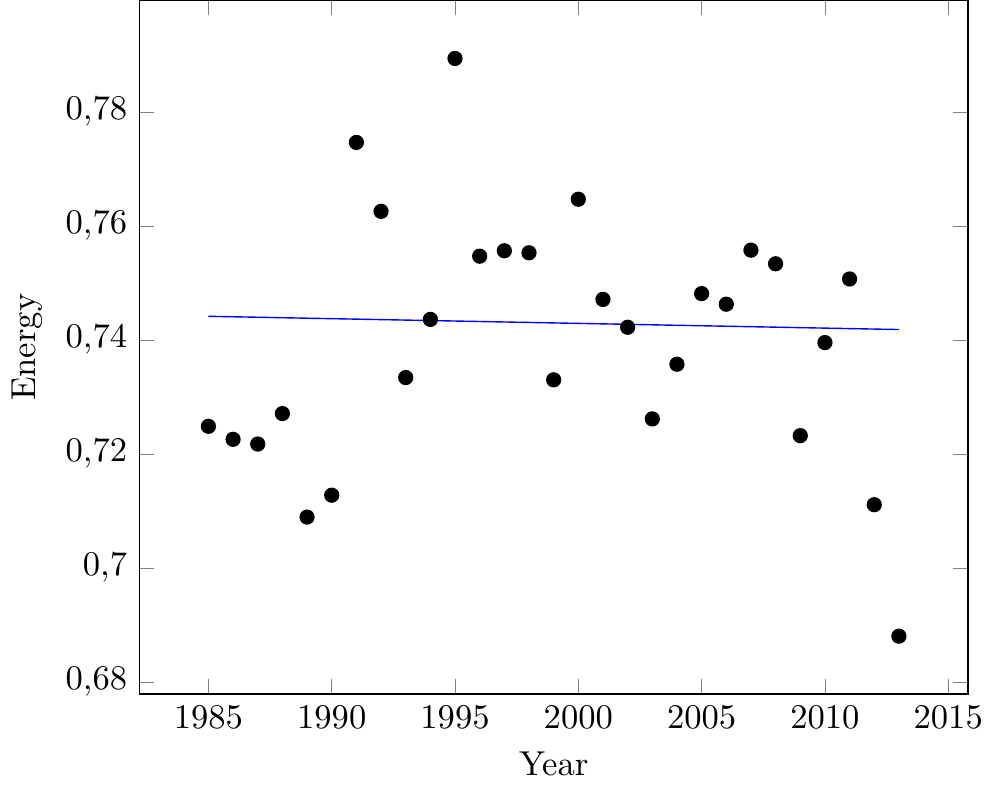}
 \label{fig:energy}
 }
\subfigure[Danceability]{
\includegraphics[scale=0.5]{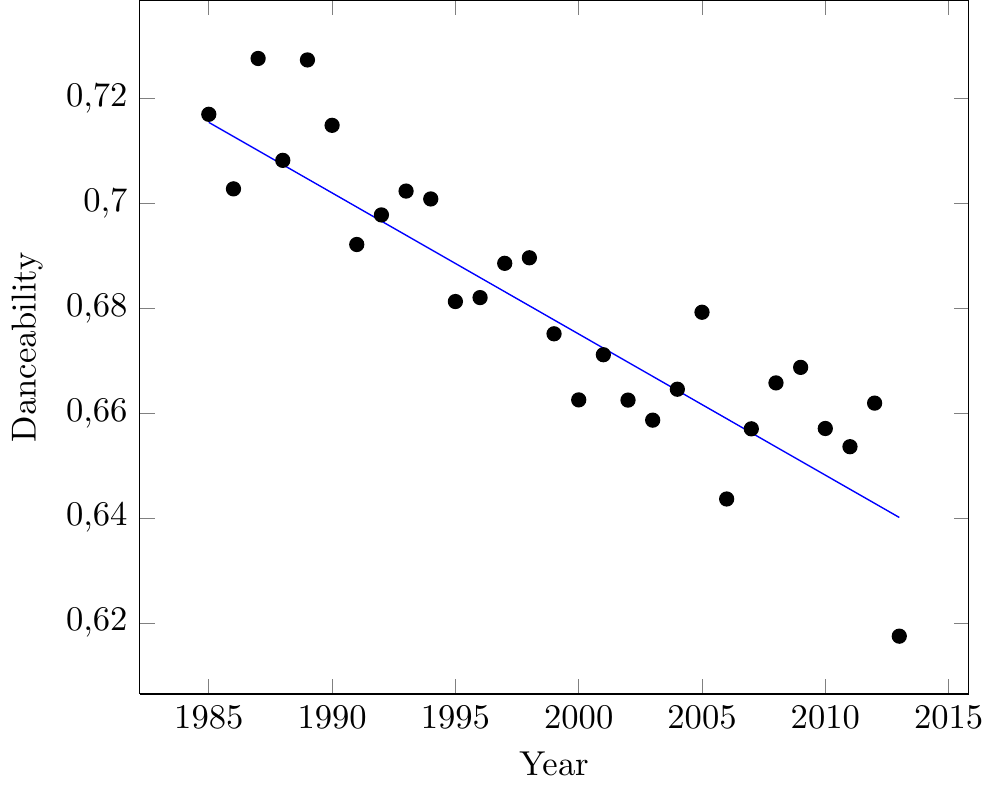}}}
% }\subfigure[Beats Differences (mean)]{   
% %  % \frametitle{}
%     \includegraphics[scale=0.5]{./plot7.pdf}
% }}
% \subfigure[Beats Differences (var)]{
% %  % \frametitle{}
%     \includegraphics[scale=0.6]{./plot8.pdf}
% }
% \subfigure[Beats Differences (skewness)]{
% %  % \frametitle{}
%     \includegraphics[scale=0.6]{./plot10.pdf}
% }
% \subfigure[Beats Differences (kurtosis)]{
% %  % \frametitle{}
%     \includegraphics[scale=0.6]{./plot9.pdf}
% }
\caption{Evolution over time of selected characteristics of top 10 songs.}
\label{fig:trend}
\end{figure}

Since a motion chart is a dynamic tool that should be viewed on a computer, a selection of features were extracted to more traditional 2-dimensional graphs with linear regressions (see Figure~\ref{fig:trend}). Since the OCC dataset contains 3,361 unique songs, the selected features from these songs were averaged per year in order to limit the amount of data points on the graph. 
A rising trend can be detected for the loudness, tempo and 1st aspect of timbre (brightness). The correlation between loudness and tempo is in line with the rule proposed by~\cite{todd1992dynamics} ``The faster the louder, the softer the slower''. Not all features have an apparent relationship with time. Energy, for instance, (see Figure~\ref{fig:energy}), doesn't seem to be correlated with time. It is also remarkable that the danceability feature computed by The Echo Nest decreases over time for dance hits. Since no detailed formula was given by The Echo Nest for danceability, this trend cannot be explained. 
  
The next sections describes an experiment which compares several hit prediction models built in this research.

\section{Dance Hit Prediction}

In this section the experimental setup and preprocessing techniques are described for the classification models built in Section~\ref{sec:techn}.

\subsection{Experiment Setup}

\begin{figure}[h]
% \centering
 \hspace{-1cm}\includegraphics[scale=0.55]{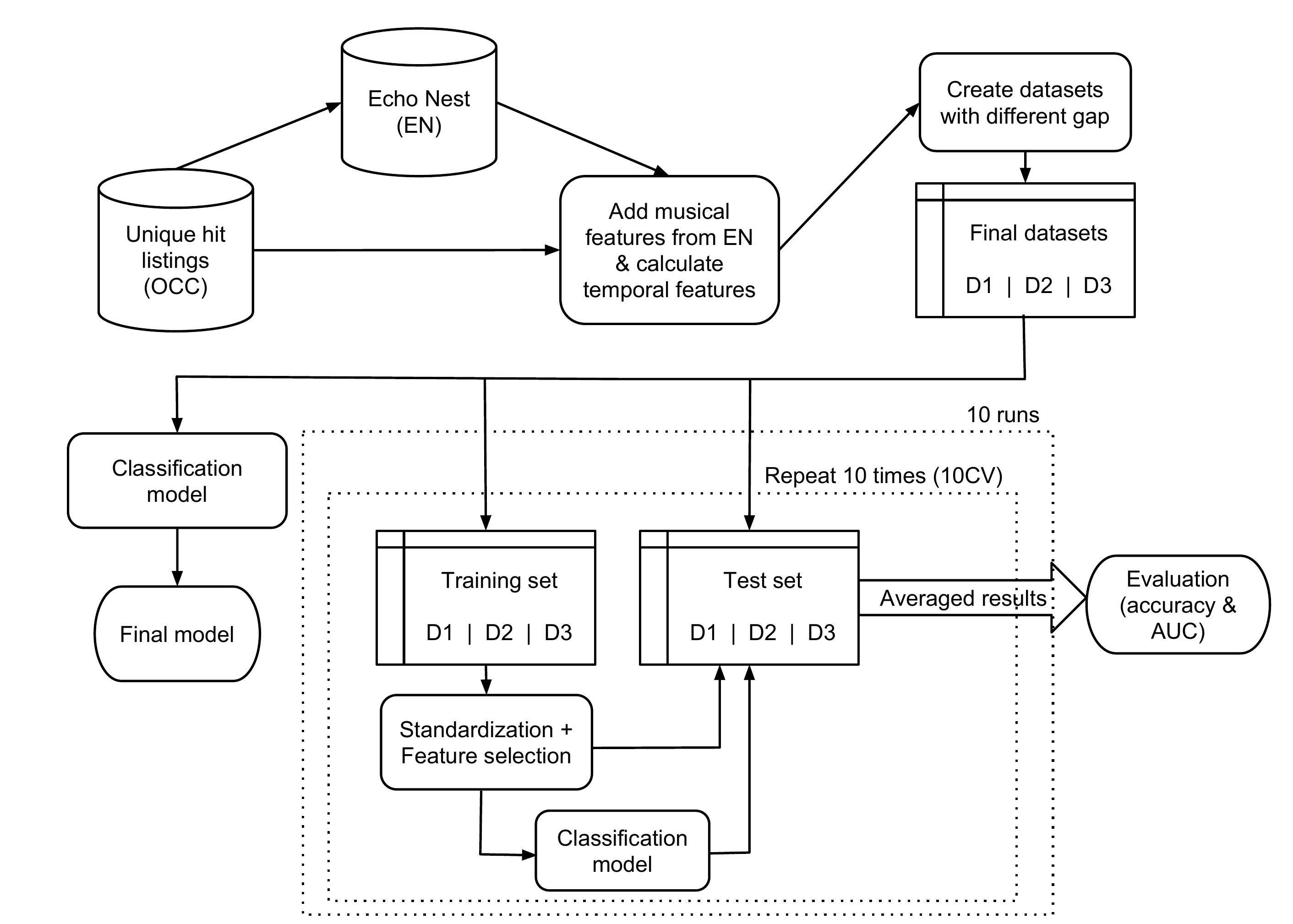}
 \caption{Flow chart of the experimental setup.}
 \label{fig:flow}
\end{figure}

Figure~\ref{fig:flow} shows a graphical representation of the setup of the experiment described in Section~\ref{sec:exp}. The dataset used for the hit prediction models in this section is based on the OCC listings. The reason for this is that this data contains top 40 songs, not just top 10. This will allow us to create a ``gap'' between the two classes. Since the previous section showed that the characteristics of hit songs evolve over time it is not representable to use data from 1985 for predicting contemporary hits. The dataset used for building the prediction models consists of dance hit songs from 2009 until 2013.

The peak chart position of each song was used to determine if they are a dance hit or not. Three datasets were made with each a different gap between the two classes (see Table~\ref{tab:datasets}). In the first dataset (D1), hits are considered to be songs with a peak position in the top 10. Non-hits are those that only reached a position between 30 and 40.  In the second dataset (D2), the gap between hits and non-hits is smaller, as songs reaching a top position of 20 are still considered to be non-hits. Finally, the original dataset is split in two at position 20, without a gap to form the third dataset (D3).  The reason for not comparing a top 10 hit with a song that did not appear in the charts is to avoid doing accidental genre classification. If a hit dance song would be compared to a song that does not occur in the hit listings, a second classification model would be needed to ensure that this non-hit song is in fact a dance song. If not, the developed model might distinguish songs based on whether or not they are a dance song instead of a hit. However, it should be noted that not all songs on the dance hit lists are in fact the same type of dance songs, there might be subgenres. Still, they will probably share more common attributes than songs from a random style, thus reducing the noise in the hit classification model.
The sizes of the three datasets are listed in Table~\ref{tab:datasets}, the difference in size can be explained by the fact that songs are excluded in D1 and D2 to form the gap. In the next sections, models are built and compare the performance of classifiers on these three datasets.

\begin{table}[h]
\centering
\caption{Datasets used for the dance hit prediction model.}
 \begin{tabular}{l|ccc}
 \toprule
 Dataset & Hits & Non-hits & Size\\
 \midrule
 D1 & Top 10 & Top 30-40 & 400\\
 D2 & Top 10 & Top 20-40 & 550\\
 D3 & Top 20 & Top 20-40 & 697\\
 \bottomrule 
 \end{tabular}
 \label{tab:datasets}
\end{table}

The Open Source software Weka was used to create the models~\citep{witten2005data}. Weka's toolbox and framework is recognized as a landmark system in the data mining and machine learning field~\citep{hall2009weka}. 

\subsection{Preprocessing}
\label{sec:pre}
\begin{figure}[h]
\centerline{
\begin{tikzpicture}
 \begin{axis}[
  ybar=8pt,%=8pt, % configures ‘bar shift’
     enlargelimits=0.25,
     ylabel={Number of instances},
 legend pos=outer north east,
 symbolic x coords={D1, D2, D3},
 % symbolic x coords={1-10vs20-40, 1-10vs30-40 ,1-20vs20-40},
 xtick=data,
 nodes near coords,
 ]
 \addplot[fill=transfertoclient] coordinates {(D2, 253) (D1,253)(D3,400)};\addlegendentry{Hits}
 \addplot[fill=rendering] coordinates {(D2, 297)(D1,147)(D3,297)};\addlegendentry{Non-hits}
 \end{axis}
 \end{tikzpicture}}
 \caption{Class distribution.}
 \label{fig:balance}
\end{figure}
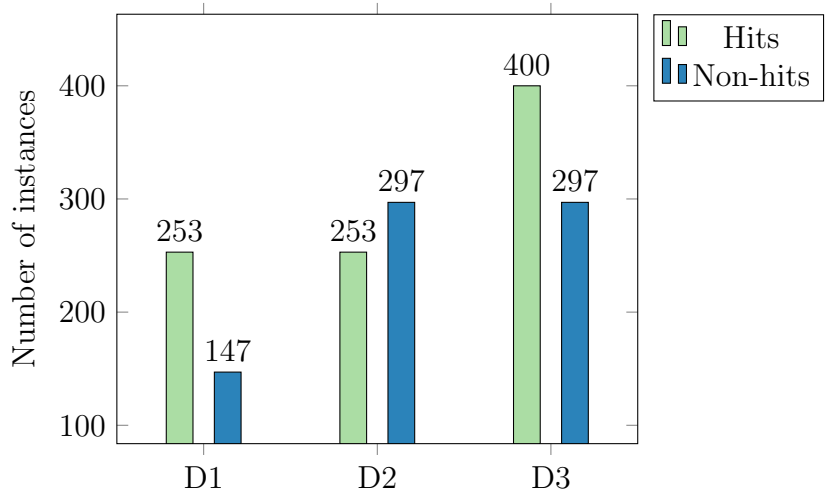

The class distribution of the three datasets used in the experiment is displayed in Figure~\ref{fig:balance}. Although the distribution is not heavily skewed, it is not completely balanced either. Because of this the use of the accuracy measure to evaluate our results is not suited and the area under the receiver operating curve (AUC)~\citep{fawcett2004roc} was used instead (see section \ref{sec:auc}).

All of the features in the datasets were standardized using statistical normalization and feature selection was done (see Figure~\ref{fig:flow}), using the procedure CfsSubsetEval from Weka with GeneticSearch. This procedure uses the individual predictive ability of each feature and the degree of redundancy between them to evaluate the worth of a subset of features~\citep{hall1999correlation}. Feature selection was done in order to avoid the ``curse of dimensionality'' by having a very sparse feature set.~\cite{mckay2006jsymbolic} point to the fact that having a limited amount of features allows for a thorough testing of the model with limited instances and can thus improve the quality of the classification model. %The reason for this is that the required number of labelled training and testing samples increases exponentially with the number of features. 
Added benefits are the improved comprehensibility of a model with a limited amount of highly predictive variables~\citep{hall1999correlation} and better performance of the learning algorithm~\citep{piramuthu2004evaluating}.
%    $x_n = \frac{x - \mu}{\sigma}$

The feature selection procedure in Weka reduces the data to 35--50 attributes, depending on the dataset. The most commonly occurring features after feature selection are listed in Table~\ref{tab:feat}. %The selected features were also stable across the folds. 
Interesting to note is that the features \emph{danceability} and \emph{energy} both disappear from the reduced datasets, except for danceability which stays in the D3 dataset. This could be explained by the fact that these features are calculated by The Echo Nest based on other features. 

\begin{table}
\caption{The most commonly occurring features in D1, D2 and D3 after FS.}
\label{tab:feat}
 \begin{tabular}{lc|lc}
 \toprule
  Feature & Occurance & Feature & Occurance \\
\midrule
 Beatdiff (range) 	&	3	&	 Timbre 1 (mean) 	&	2	\\
 Timbre 1 (80 perc) 	&	3	&	 Timbre 1 (median) 	&	2	\\
 Timbre 1 (max) 	&	3	&	 Timbre 2 (max) 	&	2	\\
 Timbre 1 (stdev) 	&	3	&	 Timbre 2 (mean) 	&	2	\\
 Timbre 2 (80 perc) 	&	3	&	 Timbre 2 (range) 	&	2	\\
 Timbre 3 (mean)	&	3	&	 Timbre 3 (var) 	&	2	\\
 Timbre 3 (median) 	&	3	&	 Timbre 4 (80 perc) 	&	2	\\
 Timbre 3 (min) 	&	3	&	 Timbre 5 (mean) 	&	2	\\
 Timbre 3 (stdev) 	&	3	&	 Timbre 5 (stdev) 	&	2	\\
 Beatdiff (80 perc) 	&	2	&	 Timbre 6 (median) 	&	2	\\
 Beatdiff (stdev) 	&	2	&	 Timbre 6 (range) 	&	2	\\
 Beatdiff (var) 	&	2	&	 Timbre 6 (var) 	&	2	\\
Timbre 11 (80 perc) 	&	2	&	 Timbre 7 (var) 	&	2	\\
 Timbre 11 (var) 	&	2	&	 Timbre 8 (Median) 	&	2	\\
 Timbre 12 (kurtosis) 	&	2	&	 Timbre 9 (kurtosis) 	&	2	\\
 Timbre 12 (Median) 	&	2	&	 Timbre 9 (max) 	&	2	\\
 Timbre 12 (min) 	&	2	&	 Timbre 9 (Median) 	&	2	\\

 \bottomrule
 \end{tabular}
 \end{table}

\section{Classification Techniques}
\label{sec:techn}

A total of five models were built for each dataset using diverse classification techniques. The two first models (decision tree and ruleset) can be considered as the easiest to understand classification models due to their linguistic nature~\citep{martensSIGKDD08}. The other three models focus on accurate prediction. In the following subsections, the individual algorithms are briefly discussed together with their main parameters and settings, followed by a comparison in Section~\ref{sec:auc}.
The AUC values mentioned in this section are based on 10-fold cross validation performance \citep{witten2005data}. The shown models are built on the entire dataset.

\subsection{C4.5 Tree}

A decision tree for dance hit prediction was built with J48, Weka's implementation of the popular C4.5 algorithm~\citep{witten2005data}.
  
The tree data structure consists of decision nodes and leaves. The class value is specified by the leaves, in this case hit or non-hit, and the nodes specify a test of one of the features. When a path from the node to a leave is followed based on the feature values of a particular song, a predictive rule can be derived~\citep{ruggieri2002efficient}.

% Decision Trees are among the most widely used methods for inductive inference.
% The hypothesis is represented using a decision tree.

A ``divide and conquer'' approach is used by the C4.5 algorithm to build trees recursively~\citep{quinlan1993}. This is a top down approach, in which a feature is sought that best separates the classes, followed by pruning of the tree~\citep{wu2008top}. This pruning is performed by a subtree raising operation in an inner cross-validation loop (3 folds by default in Weka)~\citep{witten2005data}.   %subtree raising

Decision trees have been used in a broad range of fields such as credit scoring~\citep{hand1997statistical}, land cover mapping~\citep{friedl1997decision}, medical diagnosis~\citep{wolberg1990multisurface},  estimation of toxic hazards~\citep{cramer1976estimation},  predicting customer behaviour changes~\citep{kim2005detecting} and others. 

\begin{figure}[h]
 \centering
 \includegraphics{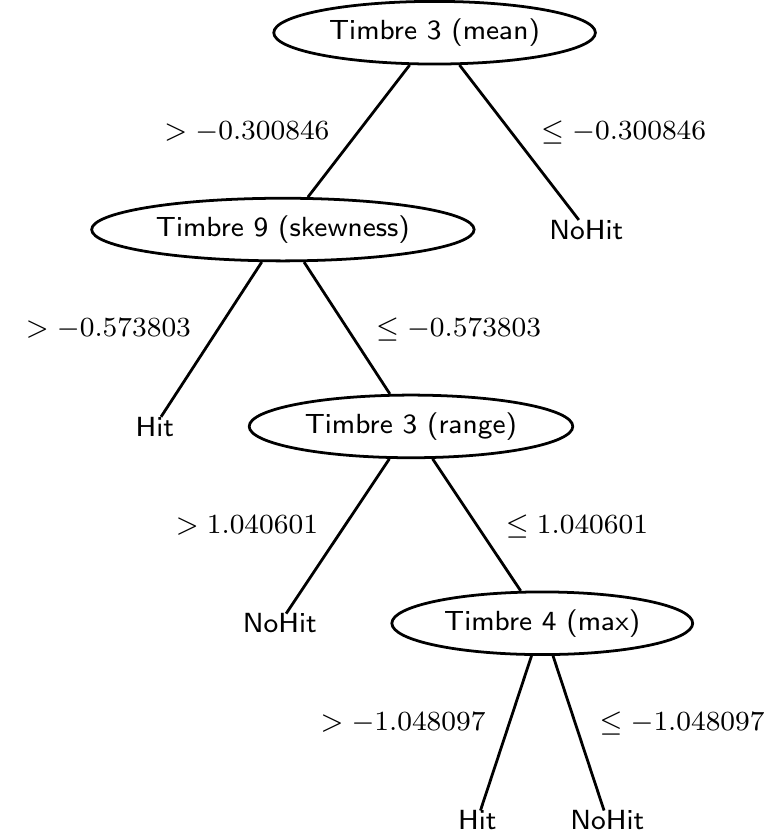}%[clip=true,bb=60 530 595 800]
 % tree.pdf: 595x842 pixel, 72dpi, 20.99x29.70 cm, bb=0 0 595 842
 \caption{C4.5 decision tree.}
 \label{fig:tree}
\end{figure}

For the comparative tests in Section~\ref{sec:auc} Weka's default settings were kept for J48. In order to create a simple abstracted model on dataset D1 (FS) for visual insight in the important features, a less accurate model (AUC 0.54) was created by pruning the tree to depth four.%with a lower confidence factor (0.01) and a larger minimum number of objects (10). 
The resulting tree is displayed in Figure~\ref{fig:tree}. It is noticeable that time differences between the third, fourth and ninth timbre vector seem to be important features for classification.

% %     a   b   <-- classified as
 %  207  46 |   a = A
 %   98  49 |   b = B
%  Confidence factor = 0.01, minNumObj = 10, AUC: 0.587%,  Accuracy: 64\%

\subsection{RIPPER Ruleset}

Much like trees, rulesets are a useful tool to gain insight in the data. They have been used in other fields to gain insight in diagnosis of technical processes~\citep{isermann1997trends}, credit scoring~\citep{baesens2003using}, medical diagnosis~\citep{kononenko2001machine}, customer relationship management~\citep{ngai2009application} and more. 

In this section JRip, Weka's implementation of the propositional rule learner RIPPER~\citep{Cohen95}, was used to inductively build ``if-then'' rules. The ``Repeated Incremental Pruning to Produce Error Reduction algorithm'' (RIPPER), uses sequential covering to generate the ruleset. In a first step of this algorithm, one rule is learned and the training instances that are covered by this rule are removed. This process is then repeated~\citep{hall2009weka}.

\begin{table}[h]
  \small
  \centering
  \caption{RIPPER ruleset.}
  \begin{tabular}{l}
  \midrule
 (T1mean $\le$ -0.020016) and (T3min $\le$ -0.534123) and (T2max $\ge$ -0.250608) $\Rightarrow$ NoHit\\
(T880perc $\le$ -0.405264) and (T3mean $\le$ -0.075106) $\Rightarrow$ NoHit\\
$\Rightarrow$ Hit\\
 %     a   b   <-- classified as
 %  203  50 |   a = A
 %   88  59 |   b = B
\bottomrule
\end{tabular}
 \label{tab:rules}
 \end{table}
 
    %Accuracy: 65.5\%

% 
%      
%      $\rightarrow$ default numNo (2), folds (3) %min totla weight of instances in a rule)
%    

The ruleset displayed in Table~\ref{tab:rules} was generated with Weka's default parameters for number of data instances (2) and folds (3) (AUC = 0.56 on dataset D1, see Table~\ref{tab:results}). It's notable that the third timbre vector is an important feature again. It would appear that this feature should not be underestimated when composing dance songs.

% xxx important timbre features - grey 66 phd wessel

\subsection{Naive Bayes}

    The naive Bayes classifier estimates the probability of a hit or non-hit based on the assumption that the features are conditionally independent. This conditional independence assumption is represented by equation (\ref{eq:ind}) given class label $y$~\citep{tan2007introduction}.
    
    \begin{equation}
P(\mathbf{x}|Y = y) = \prod_{j=1}^M P(x_j|Y=y),
\label{eq:ind}
\end{equation}
    whereby each attribute set $\mathbf{x}=\{x_1,x_2,\dots,x_N\}$ consists of $M$ attributes. 
    
    Because of the conditional dependence assumption, the class-conditional probability for every combination of $\mathbf{X}$ does not need to be calculated. Only the conditional probability of each $x_i$ given Y has to be estimated. This offers a practical advantage since a good estimate of the probability can be obtained without the need for a very large training set.

    Naive Bayes classifies a test record by calculating the posterior probability for each class Y~\citep{lewis1998naive}:
    
    \begin{equation}
P(\mathbf{Y}|x) = \frac{P(Y)\cdot \prod_{j=1}^M P(x_j|Y)}{P(\mathbf{x})}
\label{eq:nb}
\end{equation}

    Although this independence assumption is generally a poor assumption in practice, numerous studies prove that naive Bayes competes well with more sophisticated classifiers~\citep{rish2001empirical}. In particular, naive Bayes seems to be particularly resistant to isolated noise points, robust to irrelevant attributes, but its performance can degrade by correlated attributes~\citep{tan2007introduction}. Table~\ref{tab:results} confirms that Naive Bayes performs very well, with an AUC of 0.65 on dataset D1 (FS).

    \subsection{Logistic Regression}
%9 seed for the percentage split!
The SimpleLogistic function in Weka was used to build a logistic regression model~\citep{witten2005data}. %with the default parameters and the number of boosting iterations set to zero . 

%Weka's SimpleLogistic function was used to build a logistic regression model, which was fitted using LogitBoost. The LogitBoost algorithm performs additive logistic regression . Boosting algorithms like LogitBoost sequentially apply a classification algorithm, a simple regression function in this case, to reweighted versions of training data. For many classifiers the simple boosting strategy results in dramatic performance improvements~\citep{friedman2000additive}.

%    $\rightarrow$ default numBoostingIterations (0)

\begin{figure}[h]
\centering
\includegraphics{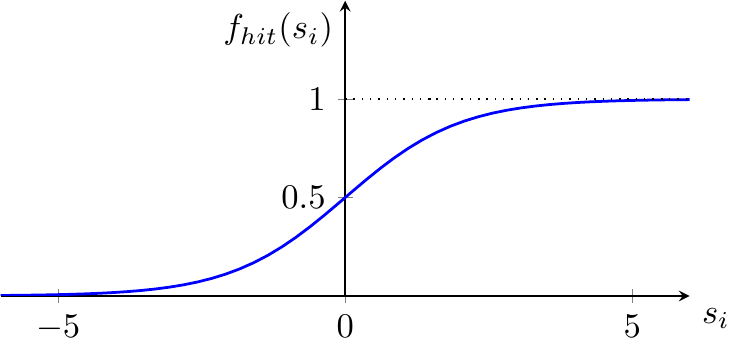}
\caption{Probability that song $i$ is a dance hit.}
\label{fig:log}
\end{figure}

Equation (\ref{eq:lr}) shows the output of a logistic regression, whereby $f_{hit}(s_{i})$ represents the probability that a song $i$ with $M$ features $x_j$ is a dance hit. This probability follows a logistic curve, as can be seen in Figure~\ref{fig:log}. The cut-off point of 0.5 will determine if a song is classified as a hit or a non-hit. With AUC = 0.65 for dataset D1 and AUC=0.67 for dataset D2 (see Table~\ref{tab:results}), logistic regression performs best for this particular classification problem.

\begin{equation}
        f_{hit}(s_{i}) = \frac{1}{1+e^{-s_{i}}}  \hspace{1cm}\text{  whereby  } s_{i} = b + \sum_{j=1}^{M} a_j \cdot x_j
\label{eq:lr}
\end{equation}

   Logistic regression models generally require limited computing power and are less prone to overfitting than other models such as neural networks~\citep{tu1996advantages}. Like the previously mentioned models, they are also used in a number of domains, such as the creation of habitat models for animals~\citep{pearce2000evaluating}, medical diagnosis~\citep{kurt2008comparing}, credit scoring~\citep{wiginton1980note} and others.
   
%In the previous sections, two comprehensible models are developed. These models provide crisp classification, which means that they determine if a musical piece is either composed by a certain composer or not. They do not offer a continuous measure that indicates ``how much'' characteristics of a certain composer are in a piece. In this section, a scoring model is developed that can accurately describe \emph{how well} a musical piece belongs to a composer's style. 
% CRISP classification~\citep{karayiannis1994fuzzy} 

% They then take the majority vote of the sequence of classifiers produced.
% Goal is a predictive model that gives an indication of how big the chance is that a composer belongs to a certain category. 
% Classifier for building linear logistic regression models. 

% LogitBoost with simple regression functions as base learners is used for fitting the logistic models. The optimal number of LogitBoost iterations to perform is cross-validated, which leads to automatic attribute selection. 
%~\citep{sumner2005speeding}

% Builds a Logistic Regression model, fitting them using LogicBoost with simple regression functions as base learners and determining how many iteration to perform using cross-validation. 

\subsection{Support Vector Machines}

%     Logistic:  multinomial logistic regression (ridge estimator)
%      $\rightarrow$ default ridge (1.0E-8), maxIts (-1)

% \section{The Support Vector Machine}\label{sec:svm}

Weka's sequential minimal optimization algorithm (SMO) was used to build two support vector machine classifiers. The support vector machine (SVM) is a learning procedure based on the statistical learning theory~\citep{vapnik}.
Given a training set of $N$ data points $\{({\bf x}_{i}, y_{i})\}_{i=1}^{N}$ with input data ${\bf x}_{i} \in \bkRrm^{n}$ and
corresponding binary class labels $y_{i} \in \{-1,+1\}$, the SVM classifier should fulfill following  conditions.~\citep{cri2000,vapnik}: \be
\label{baseconstr}
\left\{
\begin{array}{lcr}
{\bf w}^{T} \varphibold({\bf x}_{i}) + b \geq +1, & \,\,\,\, & {\rm if} \,\, \,\, y_{i} = +1 \\
{\bf w}^{T} \varphibold({\bf x}_{i}) + b \leq -1, & \,\,\,\, &
{\rm if} \,\, \,\, y_{i} = -1
\end{array}
\right.
\ee
which is equivalent to

\begin{equation} \textstyle
y_{i} [{\bf w}^{T} \varphibold({\bf x}_{i}) + b] \geq 1,
\hspace*{5mm} i = 1, ... , N.
\end{equation}

The non-linear function $\varphibold(\cdot)$ maps the input space
to a high (possibly infinite) dimensional feature space. In this
feature space, the above inequalities basically construct a
hyperplane ${\bf w}^T\varphibold({\bf x})+b=0$ discriminating
between the two classes. By minimizing  ${\bf w}^{T} {\bf w}$, the margin between both classes is maximized.

\begin{figure}[tb]
\centering
\includegraphics[scale=1.4]{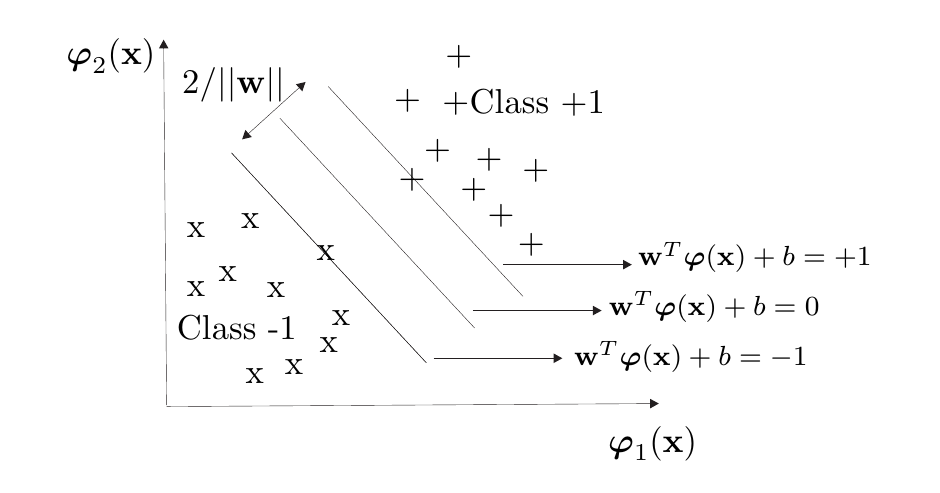}
% \begin{psfrags}
% \psfrag{x}[][]{x}
% \psfrag{+}[][]{+}
% \psfrag{E1}[][]{\ \ \ \footnotesize{${\bf w}^T\varphibold({\bf x})+b=-1$}}
% \psfrag{E2}[][]{\ \ \ \ \footnotesize{${\bf w}^T\varphibold({\bf x})+b=0$}}
% \psfrag{E3}[][]{\ \ \ \ \footnotesize{${\bf w}^T\varphibold({\bf x})+b=+1$}}
% \psfrag{E4}[][]{$\varphibold_1({\bf x})$}
% \psfrag{E5}[][]{$\varphibold_2({\bf x})$}
% \psfrag{E6}[][]{$2/||{\bf w}||$}
% \psfrag{E7}[][]{Class +1}
% \psfrag{E8}[][]{Class -1}
% \includegraphics[width = 0.54\textwidth]{SVM2.eps}
%   \end{psfrags}
\caption{Illustration of SVM optimization of the margin in the feature space.}
\label{fig:svm}
\end{figure}

In primal weight space the classifier then takes the form
\begin{equation} \label{eq:primclass} \textstyle
y({\bf x}) = {\rm sign}[{\bf w}^{T} \varphibold({\bf x}) + b] ,
\end{equation}
but, on the other hand, is never evaluated in this form. One
defines the convex optimization problem:
\begin{equation} \label{eq:V1} \textstyle
\min_{{\bf w}, b, \xibold } \mathcal{J}({\bf w}, b, \xibold) =
\frac{1}{2} {\bf w}^{T} {\bf w} + C \, \sum_{i=1}^{N} \xi_{i}
\end{equation}
subject to
\begin{equation} \label{eq:V2} \textstyle
\left\{
\begin{array}{ll}
y_{i} [{\bf w}^{T} \varphibold({\bf x}_{i}) + b] \geq 1 - \xi_{i}
,
\hspace*{1mm} &  i = 1, ... , N \\
\xi_{i} \geq 0 , \hspace*{1mm} &  i = 1, ... , N .
\end{array}
\right.
\end{equation}
The variables $\xi_{i}$ are slack variables which are needed to allow misclassifications in the set of inequalities
(e.g.,~due to overlapping distributions). The first part of the
objective function tries to maximize the margin between both
classes in the feature space and is a regularisation mechanism that penalizes for large weights, whereas the second part minimizes
the misclassification error. The positive real constant $C$ is the regularisation coefficient and should
be considered as a tuning parameter in the algorithm.

This leads to the following classifier~\citep{cri2000}: \be \label{classifier} \textstyle
y({\bf x}) = {\rm sign}[\sum_{i=1}^{N} \alpha_{i} \, y_{i} \,
K({\bf x}_{i},{\bf x}) + b], \ee whereby $K({\bf x}_i,{\bf x})=
\varphibold({\bf x}_i)^T \varphibold({\bf x})$ is taken with a
positive definite kernel satisfying the Mercer theorem.
The Lagrange multipliers $\alpha_{i}$ are then determined by optimizing the dual dual problem. The following kernel functions $K(\cdot,\cdot)$ were used:
\begin{eqnarray*}
\label{KF}
\begin{array} {ll}
% K({\bf x}, {\bf x}_{i}) = {\bf x}_{i}^{T} {\bf x}, & {\mathrm{(linear \ kernel)}} \\
K({\bf x}, {\bf x}_{i}) = (1+{{\bf x}_{i}^{T} {\bf x}}/{c})^{d}, & {\mathrm{(polynomial \ kernel)}} \\
K({\bf x}, {\bf x}_{i}) = \exp\{ - \| {\bf x} - {\bf x}_{i} \|_{2}^{2}/ \sigma^{2} \}, &  {\mathrm{(RBF \ kernel)}} \\
% K({\bf x}, {\bf x}_{i}) = \tanh( \kappa \, {\bf x}_{i}^{T} {\bf x}
% + \theta ), &
% {\mathrm{(MLP \ kernel)}}, \\
\end{array}
\end{eqnarray*}
where $d$, $c$ and $\sigma$ are constants. %Note that for the MLP kernel, the Mercer condition is not always satisfied.$\kappa$ and $\theta$ 

For low-noise problems, many of the $\alpha_i$ will typically be equal to zero (sparseness property).  The training observations corresponding to non-zero $\alpha_i$ are called support vectors and are located  close to the decision boundary.

As equation (\ref{classifier}) shows, the SVM classifier with non-linear kernel is a complex, non-linear function. Trying to comprehend the logics of the classifications made is quite difficult, if not impossible~\citep{martensALBA2008, martensprovost13EDC}.
   
In this research, the Polynomial kernel and RBF kernel were used to build the models. %When building the model and extra step is performed to normalize the training and test sets.
Although Weka's default settings were used in the previous models, the hyperparameters for the SVM model were optimized. To determine the optimal settings for the regularisation parameter $C$ (1, 3, 5,\dots 21), the $\sigma$ for the RBF kernel ($\frac{1}{\sigma^2}$ = 0.00001, 0.0001,\dots 10) and the exponent $d$ for the polynomial kernel (1,2), GridSearch was used in Weka. The choice of hyperparameters to test was inspired by settings suggesting by~\cite{wikiweka2013}. GridSearch performs 2-fold cross validation on the initial grid. This grid is determined by the two input parameters ($C$ and $\sigma$ for the RBF kernel, $C$ and $d$ for the polynomial kernel). 10-fold cross validation is then performed on the best point of the grid based on the weighted AUC by class size and its adjacent points. If a better pair is found, the procedure is repeated on its neighbours until no better pair is found or the border of the grid is reached~\citep{grid2013}. This hyperparameter optimization is performed in the ``classification model'' box in Figure~\ref{fig:flow}. The resulting AUC-value is 0.59 for the SVM with polynomial and 0.56 for the SVM with RBF kernel on D1 (FS) (see Table~\ref{tab:results}). 

% $  e^-(gamma * <x-y, x-y>^2)$--optimizing sigma here?
%   CVParameterSelection
%    GridSearch: 
%    Multiple parameters:
%    c: 1--21 (+2)
%    gamma (RBF): 0.00001--10 (*10)
%   exponent (Poly): 0.1--1,6 (+0.5)

\section{Results}
\label{sec:auc}

In this section, two experiments are described. The first one builds models for all of the datasets (D1, D2 \& D3), both with and without feature selection. The evaluation is done by taking the average of 10 runs, each with a 10-fold cross validation procedure. In the second experiment, the performance of the classifiers on the best dataset is compared with an out-of-time test set.

\subsection{Full Experiment With Cross-validation}
\label{sec:exp}
A comparison of the accuracy and the AUC is displayed in Table~\ref{tab:aresults} and~\ref{tab:results} for all of the above mentioned classifiers. The tests were run 10 times, each time with stratified 10-fold cross validation (10CV), both with and without feature selection (FS). This process is depicted in Figure~\ref{fig:flow}. As mentioned in Section~\ref{sec:pre}, AUC is a more suited measure since the datasets are not entirely balanced~\citep{fawcett2004roc}, yet both are displayed to be complete. During the cross validation procedure, the dataset is divided into 10 folds. 9 of them are used for model building and 1 for testing. This procedure is repeated 10 times. The displayed AUC and accuracy in this subsection are the average results over the 10 test sets and the 10 runs. The resulting model is built on the entire dataset and can be expected to have a performance which is at least as good as the 10CV performance. A total of 10 runs were performed with the 10CV prodedure and the average results are displayed in Table~\ref{tab:aresults} and~\ref{tab:results}.
A Wilcoxon signed-rank test is conducted to compare the performance of the models with the best performing model. The null hypothesis of this test states: ``There is no difference in the performance of a model with the best model''.
% omparing two related samples, matched samples, or repeated measurements on a single sample to assess whether their population mean ranks differ
%TO FIX XXX One minor remark though: could the authors clarify in their camera-ready version the meaning of the p-values below tables 6-8? These are not entirely clear to me, and the explanation in the text does not clearly say what the null hypothesis and the test statistic are. It seems strange that all the best scores have a p-value larger than 0.05.
%  test a hypothesis about the location (median) of a population distribution. It often involves the use of matched pairs, for example, before and after data, in which case it tests for a median difference of zero.

% The Wilcoxon Signed Ranks test does not require the assumption that the population is normally distributed.

\begin{table}
\centering
 \caption{Results with 10-fold validation (accuracy).}
\begin{tabular}{l|cc|cc|cc}
 \toprule
  Accuracy (\%)                    &\multicolumn{2}{c}{D1} &\multicolumn{2}{c}{D2}&\multicolumn{2}{c}{D3} \\
 {} & - & FS & - & FS & - & FS\\
 \midrule	
C4.5&	\textit{57.05}	&	\textit{58.25}	&	\textit{54.95}	&	\textit{54.67	}&	\textit{54.58	}&	\textit{54.74	}\\
RIPPER&	\textit{60.95}	&	\textit{62.43}	&	\textit{56.69}	&	\textit{56.42	}&	\textit{57.18	}&	\textit{56.41	}\\
Naive Bayes&	\underline{\textbf{65}}	&	\underline{\textbf{65}}	&	\textit{60.22}	&	\textit{58.78	}&	59.57	&	\textit{59.18	}\\
Logistic regression&	\textbf{64.65}	&	\textbf{64}	&	\underline{\textbf{62.64}}	&	\textbf{60.6	}&	\textbf{60.12	}&	\textit{59.75	}\\
SVM (Polynomial)&	\textbf{64.97}	&	\textbf{64.7}	&	\textbf{61.55}	&	\underline{\textbf{61.6	}}&	\underline{\textbf{61.04	}}&	\underline{\textbf{61.07	}}\\
SVM (RBF)&	\textbf{64.7}	&	\textbf{64.63}	&	\textit{59.8}	&	\textit{59.89	}&	\textbf{60.8	}&	\textbf{60.76	}\\
 \bottomrule
 \end{tabular} 
\flushleft \hspace{0.4cm} \small FS = feature selection, $p<0.01$: italic, $p>0.05$: bold, best: \underline{bold}.
\label{tab:aresults}
  \end{table}

\begin{table}[h]
\centering
 \caption{Results for 10 runs with 10-fold validation (AUC).}
\begin{tabular}{l|cc|cc|cc}
 \toprule
AUC                  &\multicolumn{2}{c}{D1} &\multicolumn{2}{c}{D2}&\multicolumn{2}{c}{D3} \\
 {} & - & FS & - & FS & - & FS\\
 \midrule
C4.5&	\textit{0.53	}&	\textit{0.55	}&	\textit{0.55	}&	\textit{0.54	}&	\textit{0.54	}&	\textit{0.53	}\\
RIPPER&	\textit{0.55	}&	\textit{0.56	}&	\textit{0.56	}&	\textit{0.56	}&	\textit{0.54	}&	\textit{0.55	}\\
Naive Bayes&	\textbf{0.64	}&	\underline{\textbf{0.65	}}&	\textit{0.64	}&	\textbf{0.63	}&	\textbf{0.6	}&	\textit{0.61	}\\
Logistic regression&	\underline{\textbf{0.65	}}&	\underline{\textbf{0.65	}}&	\underline{\textbf{0.67	}}&	\underline{\textbf{0.64	}}&	\underline{\textbf{0.61	}}&	\underline{\textbf{0.63	}}\\
SVM (Polynomial)&	\textit{0.6	}&	\textit{0.59	}&	\textit{0.61	}&	\textit{0.61	}&	\textit{0.58	}&	\textit{0.58	}\\
SVM (RBF)&	\textit{0.56	}&	\textit{0.56	}&	\textit{0.59	}&	\textit{0.6	}&	\textit{0.57	}&	\textit{0.57	}\\
\bottomrule
 \end{tabular}
 \flushleft \hspace{1.1cm} \small FS = feature selection, $p<0.01$: italic, $p>0.05$: bold, best: \underline{bold}.
\label{tab:results}
\end{table}

% IMPORTANT TIMBRAL FEATURES. 
%This is nice because several perceptual studies have shown that timbre was well discriminated by attack, and brightness qualities, which is pretty well confirmed by the data-driven representation here. XXX

As described in the previous section, decision trees and rulesets do not always offer the most accurate classification results, but their main advantage is their comprehensibility~\citep{craven1996extracting}. It is rather surprising that support vector machines do not perform very well on this particular problem. The overall best technique seems to be the logistic regression, closely followed by naive Bayes. Another conclusion from the table is that feature selection seems to have a positive influence on the AUC for D1 and D3. As expected, the overall best results when taking into account both AUC and accuracy can be obtained using the dataset with the biggest gap, namely D1.

\begin{table}[h]
\caption{Results for 10 runs on D1 (FS) with 10-fold cross validation compared with the split test set.}
\centering
 \begin{tabular}{l|cc|cc}
 \toprule
 &  \multicolumn{2}{c}{AUC}& \multicolumn{2}{c}{accuracy (\%)}\\
 &   split & 10CV & split & 10CV \\
 % J48 & 67.0886\% & 0.597\\
 % JRip & 64.557\% & 0.586\\
 % Naive Bayes & 70.8861\% & 0.709\\
 % SimpleLogistic & 65.8228\% & 0.670\\
 % Logistic & \textbf{78.481\%} & \textbf{0.795}\\
 % SMO (RBF Kernel) & \textbf{78.481\%} & 0.736\\
 % SMO (PolyKernel) & 77.2152\% & 0.733\\
 \midrule
   C4.5 & 		0.62		& \textit{0.55}			&62.50	&\textit{58.25}\\
 RIPPER & 		0.66 		& \textit{0.56}	  		& \underline{\textbf{85}} 	&\textit{62.43}\\	
 Naive Bayes & 		0.79 		& \underline{\textbf{0.65}}	 &77.50	& \underline{\textbf{65}}\\
Logistic regression &  	\underline{\textbf{0.81}} 			& \underline{\textbf{0.65}}	& 80	&\textbf{64}\\
 SVM (Polynomial) & 	0.729 		& \textit{0.59} 		&\underline{\textbf{85}}	&\textbf{64.7}\\
 SVM (RBF) & 		0.57 		& \textit{0.56} 		&82.5	&\textbf{64.63}\\
 \bottomrule
 \end{tabular} 
 \flushleft \hspace{2cm} \small $p<0.01$: italic, $p>0.05$: bold, best: \underline{bold}.
\label{fig:results2}
\end{table}

\begin{figure}[h]
    \centering
       \includegraphics[scale=0.5]{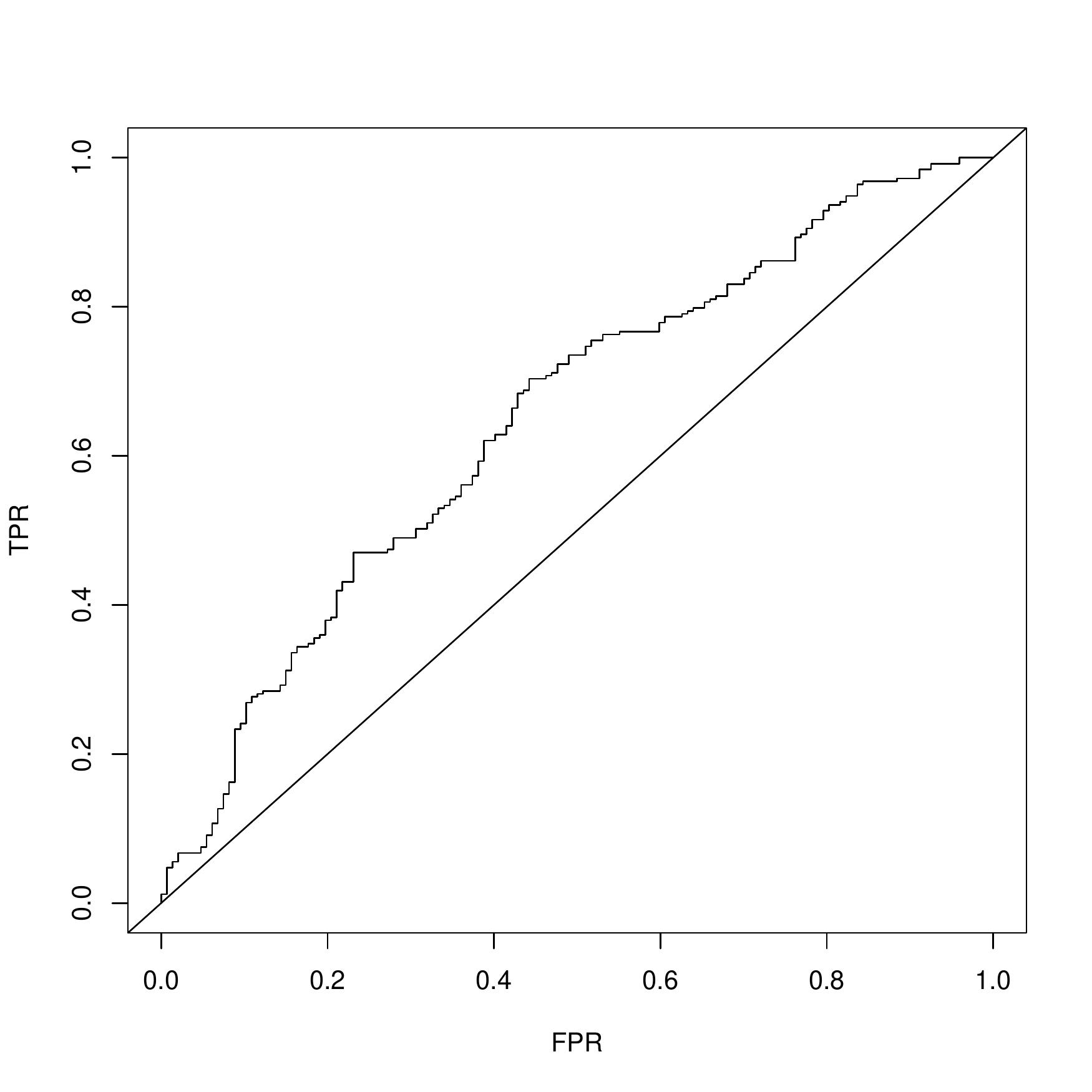}
       \caption{ROC for Logistic regression}
       \label{fig:roc}
    \end{figure}

The overall best model seems to be logistic regression. The receiver operating curve (ROC) is displayed in Figure~\ref{fig:roc}. The ROC curve displays the trade-off between true positive rate (TPR) and false negative rate (FNR) of the logistic classifier with 10-fold cross validation for D1 (FS). The model clearly scores better than a random classification, which is represented by the diagonal through the origin.

The confusion matrix of the logistic regression shows that 209 hits (i.e. 83\% of the actual hits) were accurately classified as hits and 47 non-hits classified as non-hits (i.e. 32\% of the actual non-hits). Yet overall, the model is able to make a fairly good distinction between classes, which proves that the dance hit prediction problem can be tackled as realistic top 10 versus top 30-40 classification problem with logistic regression.
  
  \begin{table}[h]
 \caption{Confusion matrix logistic regression.}
\centering
\begin{tabular}{ccl}
\hline
   a  & b &  $\leftarrow$ classified as\\
\hline
  209 & 44 &   a = hit\\
  100 & 47 &   b = non-hit\\
\hline
\label{tab:clog}
\end{tabular}
\end{table}

\subsection{Experiment With Out-of-time Test Set}
\begin{figure}
  \centering
  \begin{tikzpicture}
 \begin{axis}[
  ybar=8pt,%=8pt, % configures ‘bar shift’
   enlarge x limits=0.7,
%          enlargelimits=0.25,
% bar width=12pt,
%  x=6cm,
ylabel={Number of instances},
      ymin=0,
%       legend pos= outer north east,
 symbolic x coords={training set, test set},
 xtick=data,
  nodes near coords,
 ]
 \addplot[fill=transfertoclient] coordinates {(test set, 35) (training set, 218)};\addlegendentry{Hits}
 \addplot[fill=rendering] coordinates {(test set, 5) (training set, 142)};\addlegendentry{Non-hits}
 \end{axis}
\end{tikzpicture}
\caption{Class distribution of the split training and test sets.}
\label{fig:bal2}
\end{figure}
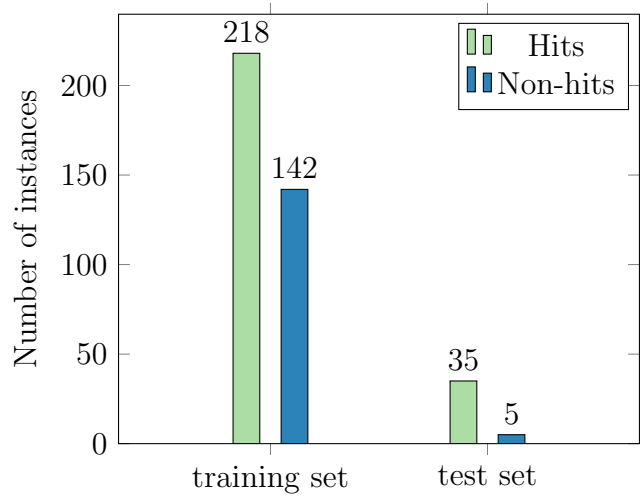

A second experiment was conducted with an out-of-time test set based on D1 with feature selection. The instances were first ordered by date, and then split into a 90\% training and 10\% test set. Table~\ref{fig:results2} confirms the good performance of the logistic regression. A peculiar observation from this table is that the model seems to be able to predict better for newer songs (AUC: 0.81 versus 0.65). This can be due to coincidence, different class distribution between training and test set (see Figure~\ref{fig:bal2}) or the structure of the dataset. One speculation of the authors is that the oldest instances of the dataset might be ``lingering'' hits, meaning that they were top 10 hits on a date before the earliest entry in the dataset, and were still present in a low position in the used hit listings. These songs would be falsely seen as non-hits, which might cause the model to predict less good for older songs. 
  
\section{Conclusion}

Multiple models were built that can successfully predict if a dance song is going to be a top 10 hit versus a lower positioned dance song. In order to do this, hit listings from two chart magazines were collected and mapped to audio features provided by The Echo Nest. Standard audio features were used, as well as more advanced features that capture the temporal aspect. This resulted in a model that could accurately predict top 10 dance hits. 

This research proves that popularity of dance songs \emph{can} be learnt from the analysis of music signals. Previous less successful results in this field speculate that their results could be due to features that are not informative enough~\citep{pachet2008hit}. The positive results from this paper could indeed be due to the use of more advanced temporal features. A second cause might be the use of ``recent'' songs only, which eliminates the fact that hit music evolves over time. It might also be due to the nature of dance music or that by focussing on one particular style of music, any noise created by classifying hits of different genres is reduced. Finally, by comparing different classifiers that have significantly different results in performance, the best model could be selected. 

This model was implemented in an online application where users can upload their audio data and get the probability of it being a hit\footnote{\url{http://antor.ua.ac.be/dance}}. 
An interesting future expansion would be to improve the accuracy of the model by including more features such as lyrics, social network information and others. The model could also be expanded to predict hits of other musical styles.
In the line of research being done with automatic composition systems~\citep{herremans2013composing}, it is also interesting to see if the classification models from this paper could be included in an optimization function (e.g., a type of fitness function) and used to generate new dance hits or improve existing ones. 
 
%controvery hit song science a science
% works because of dance hit nature?

\section*{Acknowledgement}
This research has been partially supported by the Interuniversity Attraction Poles (IUAP) Programme initiated by the Belgian Science Policy Office (COMEX project).

\bibliographystyle{plainnat}
\bibliography{paper}

\begin{thebibliography}{67}
\providecommand{\natexlab}[1]{#1}
\providecommand{\url}[1]{\texttt{#1}}
\expandafter\ifx\csname urlstyle\endcsname\relax
  \providecommand{\doi}[1]{doi: #1}\else
  \providecommand{\doi}{doi: \begingroup \urlstyle{rm}\Url}\fi

\bibitem[Baesens et~al.(2003)Baesens, Setiono, Mues, and
  Vanthienen]{baesens2003using}
B.~Baesens, R.~Setiono, C.~Mues, and J.~Vanthienen.
\newblock Using neural network rule extraction and decision tables for
  credit-risk evaluation.
\newblock \emph{Management Science}, 49\penalty0 (3):\penalty0 312--329, 2003.

\bibitem[Bertin-Mahieux et~al.(2011)Bertin-Mahieux, Ellis, Whitman, and
  Lamere]{BertinMahieux2011}
T.~Bertin-Mahieux, D.P.W. Ellis, B.~Whitman, and P.~Lamere.
\newblock The million song dataset.
\newblock In \emph{{Proceedings of the 12th International Conference on Music
  Information Retrieval ({ISMIR} 2011)}}, 2011.

\bibitem[Borg and Hokkanen(2011)]{borg83}
N.~Borg and G.~Hokkanen.
\newblock What makes for a hit pop song? what makes for a pop song?
\newblock 2011.
\newblock URL
  \url{http://cs229.stanford.edu/proj2011/BorgHokkanen-WhatMakesForAHitPopSong.pdf}.

\bibitem[Braheny(2007)]{braheny2007}
J.~Braheny.
\newblock \emph{Craft and Business of Songwriting 3rd Edition (Craft \&
  Business of Songwriting)}.
\newblock F \& W Publications, 2007.

\bibitem[Casey et~al.(2008)Casey, Veltkamp, Goto, Leman, Rhodes, and
  Slaney]{casey2008content}
M.A. Casey, R.~Veltkamp, M.~Goto, M.~Leman, C.~Rhodes, and M.~Slaney.
\newblock Content-based music information retrieval: Current directions and
  future challenges.
\newblock \emph{Proceedings of the IEEE}, 96\penalty0 (4):\penalty0 668--696,
  2008.

\bibitem[Cohen(1995)]{Cohen95}
W.~Cohen.
\newblock Fast effective rule induction.
\newblock In Armand Prieditis and Stuart Russell, editors, \emph{Proceedings of
  the 12th International Conference on Machine Learning}, pages 115--123, Tahoe
  City, CA, 1995. Morgan Kaufmann Publishers.

\bibitem[Cramer et~al.(1976)Cramer, Ford, and Hall]{cramer1976estimation}
G.M. Cramer, R.A. Ford, and R.L. Hall.
\newblock Estimation of toxic hazard—a decision tree approach.
\newblock \emph{Food and cosmetics toxicology}, 16\penalty0 (3):\penalty0
  255--276, 1976.

\bibitem[Craven and Shavlik(1996)]{craven1996extracting}
M.W. Craven and J.W. Shavlik.
\newblock Extracting tree-structured representations of trained networks.
\newblock \emph{Advances in neural information processing systems}, 8:\penalty0
  24--30, 1996.

\bibitem[Cristianini and Shawe-Taylor(2000)]{cri2000}
N.~Cristianini and J.~Shawe-Taylor.
\newblock \emph{An introduction to Support Vector Machines and Other
  Kernel-Based Learning Methods}.
\newblock Cambridge University Press, New York, NY, USA, 2000.

\bibitem[Dhanaraj and Logan(2005)]{dhanaraj2005automatic}
R.~Dhanaraj and B.~Logan.
\newblock Automatic prediction of hit songs.
\newblock In \emph{Proceedings of the International Conference on Music
  Information Retrieval}, pages 488--91, 2005.

\bibitem[Downie(2003)]{downie2003music}
J.S. Downie.
\newblock Music information retrieval.
\newblock \emph{Annual review of information science and technology},
  37\penalty0 (1):\penalty0 295--340, 2003.

\bibitem[EchoNest(2013)]{echo2013}
EchoNest.
\newblock The echo nest.
\newblock 2013.
\newblock URL \url{http://echonest.com}.

\bibitem[Essid et~al.(2006)Essid, Richard, and David]{essid2006musical}
S.~Essid, G.~Richard, and B.~David.
\newblock Musical instrument recognition by pairwise classification strategies.
\newblock \emph{Audio, Speech, and Language Processing, IEEE Transactions on},
  14\penalty0 (4):\penalty0 1401--1412, 2006.

\bibitem[Fawcett(2004)]{fawcett2004roc}
T.~Fawcett.
\newblock Roc graphs: Notes and practical considerations for researchers.
\newblock \emph{Machine Learning}, 31:\penalty0 1--38, 2004.

\bibitem[Friedl and Brodley(1997)]{friedl1997decision}
M.A. Friedl and C.E. Brodley.
\newblock Decision tree classification of land cover from remotely sensed data.
\newblock \emph{Remote sensing of environment}, 61\penalty0 (3):\penalty0
  399--409, 1997.

\bibitem[Fu et~al.(2011)Fu, Lu, Ting, and Zhang]{fu2011survey}
Z.~Fu, G.~Lu, K.M. Ting, and D.~Zhang.
\newblock A survey of audio-based music classification and annotation.
\newblock \emph{Multimedia, IEEE Transactions on}, 13\penalty0 (2):\penalty0
  303--319, 2011.

\bibitem[Google(2013)]{motion2013}
Google.
\newblock Google charts - visualisation: Motion graph, 2013.
\newblock URL
  \url{https://developers.google.com/chart/interactive/docs/gallery/motionchart}.

\bibitem[Hall et~al.(2009)Hall, Frank, Holmes, Pfahringer, Reutemann, and
  Witten]{hall2009weka}
M.~Hall, E.~Frank, G.~Holmes, B.~Pfahringer, P.~Reutemann, and I.H. Witten.
\newblock The weka data mining software: an update.
\newblock \emph{ACM SIGKDD Explorations Newsletter}, 11\penalty0 (1):\penalty0
  10--18, 2009.

\bibitem[Hall(1999)]{hall1999correlation}
M.A. Hall.
\newblock \emph{Correlation-based feature selection for machine learning}.
\newblock PhD thesis, The University of Waikato, 1999.

\bibitem[Hand and Henley(1997)]{hand1997statistical}
D.J. Hand and W.E. Henley.
\newblock Statistical classification methods in consumer credit scoring: a
  review.
\newblock \emph{Journal of the Royal Statistical Society: Series A (Statistics
  in Society)}, 160\penalty0 (3):\penalty0 523--541, 1997.

\bibitem[Herremans and S{\"o}rensen(2013)]{herremans2013composing}
D~Herremans and K~S{\"o}rensen.
\newblock Composing fifth species counterpoint music with a variable
  neighborhood search algorithm.
\newblock \emph{Expert Systems with Applications}, 40\penalty0 (16):\penalty0
  6427--6437, 2013.

\bibitem[Herremans et~al.(2013)Herremans, S\"orensen, and
  Martens]{herremans2013}
D.~Herremans, K.~S\"orensen, and D.~Martens.
\newblock Classification and generation of composer specific music.
\newblock \emph{Working paper - University of Antwerp}, 2013.

\bibitem[Houston(2013)]{houstoninstant2013}
P.~Houston.
\newblock \emph{Instant jsoup How-to}.
\newblock Packt Publishing Ltd, 2013.

\bibitem[IFPI(2012)]{ifpi2012}
IFPI.
\newblock Investing in music.
\newblock Technical report, International Federation of the Phonographic
  Industry, 2012.
\newblock URL \url{http://www.ifpi.org/content/library/investing_in_music.pdf}.

\bibitem[Isermann and Balle(1997)]{isermann1997trends}
R.~Isermann and P.~Balle.
\newblock Trends in the application of model-based fault detection and
  diagnosis of technical processes.
\newblock \emph{Control engineering practice}, 5\penalty0 (5):\penalty0
  709--719, 1997.

\bibitem[Jehan(2005)]{jehan2005creating}
T.~Jehan.
\newblock \emph{Creating music by listening}.
\newblock PhD thesis, Massachusetts Institute of Technology, 2005.

\bibitem[Jehan and DesRoches(2012)]{jehan2012}
T.~Jehan and D.~DesRoches.
\newblock \emph{{EchoNest Analyzer Documentation}}, 2012.
\newblock URL
  \url{developer.echonest.com/docs/v4/_static/AnalyzeDocumentation.pdf}.

\bibitem[Jelinek(2005)]{jelinek2005some}
F.~Jelinek.
\newblock Some of my best friends are linguists.
\newblock \emph{Language resources and evaluation}, 39\penalty0 (1):\penalty0
  25--34, 2005.

\bibitem[Kim et~al.(2005)Kim, Song, Kim, and Kim]{kim2005detecting}
J.K. Kim, H.S. Song, T.S. Kim, and H.K. Kim.
\newblock Detecting the change of customer behavior based on decision tree
  analysis.
\newblock \emph{Expert Systems}, 22\penalty0 (4):\penalty0 193--205, 2005.

\bibitem[Kononenko(2001)]{kononenko2001machine}
I.~Kononenko.
\newblock Machine learning for medical diagnosis: history, state of the art and
  perspective.
\newblock \emph{Artificial Intelligence in medicine}, 23\penalty0 (1):\penalty0
  89--109, 2001.

\bibitem[Kurt et~al.(2008)Kurt, Ture, and Kurum]{kurt2008comparing}
I.~Kurt, M.~Ture, and A.T. Kurum.
\newblock Comparing performances of logistic regression, classification and
  regression tree, and neural networks for predicting coronary artery disease.
\newblock \emph{Expert Systems with Applications}, 34\penalty0 (1):\penalty0
  366--374, 2008.

\bibitem[Lamere(2013)]{jen2013}
P.~Lamere.
\newblock jen-api - a java client for the echonest.
\newblock 2013.
\newblock URL \url{http://code.google.com/p/jen-api/}.

\bibitem[Laurier et~al.(2008)Laurier, Grivolla, and
  Herrera]{laurier2008multimodal}
C.~Laurier, J.~Grivolla, and P.~Herrera.
\newblock Multimodal music mood classification using audio and lyrics.
\newblock In \emph{Machine Learning and Applications, 2008. ICMLA'08. Seventh
  International Conference on}, pages 688--693. IEEE, 2008.

\bibitem[Leikin(2008)]{molly2008}
M.A. Leikin.
\newblock \emph{How to Write a Hit Song, 5th Edition}.
\newblock Hal Leonard, 2008.

\bibitem[Lewis(1998)]{lewis1998naive}
D.D. Lewis.
\newblock Naive (bayes) at forty: The independence assumption in information
  retrieval.
\newblock In \emph{Machine learning: ECML-98}, pages 4--15. Springer, 1998.

\bibitem[Martens(2008)]{martensSIGKDD08}
D.~Martens.
\newblock Building acceptable classification models for financial engineering
  applications.
\newblock \emph{SIGKDD Explorations}, 10\penalty0 (2):\penalty0 30--31, 2008.

\bibitem[Martens and Provost(2014)]{martensprovost13EDC}
D.~Martens and F.~Provost.
\newblock Explaining data-driven document classifications.
\newblock \emph{MIS Quarterly}, 38\penalty0 (1):\penalty0 73--99, 2014.

\bibitem[Martens et~al.(2009)Martens, {Van Gestel}, and
  Baesens]{martensALBA2008}
D.~Martens, T.~{Van Gestel}, and B.~Baesens.
\newblock Decompositional rule extraction from support vector machines by
  active learning.
\newblock \emph{IEEE Transactions on Knowledge and Data Engineering},
  21\penalty0 (2):\penalty0 178--191, 2009.

\bibitem[McKay and Fujinaga(2006)]{mckay2006jsymbolic}
C.~McKay and I.~Fujinaga.
\newblock jsymbolic: A feature extractor for midi files.
\newblock In \emph{Proceedings of the International Computer Music Conference},
  pages 302--5, 2006.

\bibitem[Ngai et~al.(2009)Ngai, Xiu, and Chau]{ngai2009application}
E.W.T. Ngai, L.~Xiu, and D.C.K. Chau.
\newblock Application of data mining techniques in customer relationship
  management: A literature review and classification.
\newblock \emph{Expert Systems with Applications}, 36\penalty0 (2):\penalty0
  2592--2602, 2009.

\bibitem[Ni et~al.(2011)Ni, Santos-Rodr{\'\i}guez, McVicar, and
  De~Bie]{ni2011hit}
Y.~Ni, R.~Santos-Rodr{\'\i}guez, M.~McVicar, and T.~De~Bie.
\newblock Hit song science once again a science?
\newblock 2011.

\bibitem[Ni et~al.(2013)Ni, Santos-Rodr{\'\i}guez, McVicar, and
  De~Bie]{score2013}
Y.~Ni, R.~Santos-Rodr{\'\i}guez, M.~McVicar, and T.~De~Bie.
\newblock Score a hit - documentation.
\newblock 2013.
\newblock URL \url{http://www.scoreahit.com/Documentation}.

\bibitem[Pachet(2012)]{pachet2012hit}
F.~Pachet.
\newblock Hit song science.
\newblock \emph{Tzanetakis \& Ogihara Tao, editor, Music Data Mining}, pages
  305--326, 2012.

\bibitem[Pachet and Roy(2008)]{pachet2008hit}
F.~Pachet and P.~Roy.
\newblock Hit song science is not yet a science.
\newblock In \emph{Proc. of the 9th International Conference on Music
  Information Retrieval (ISMIR 2008)}, pages 355--360, 2008.

\bibitem[Pearce and Ferrier(2000)]{pearce2000evaluating}
J.~Pearce and S.~Ferrier.
\newblock Evaluating the predictive performance of habitat models developed
  using logistic regression.
\newblock \emph{Ecological modelling}, 133\penalty0 (3):\penalty0 225--245,
  2000.

\bibitem[Perricone(2000)]{jack2000}
J.~Perricone.
\newblock \emph{Melody in Songwriting: Tools and Techniques for Writing Hit
  Songs (Berklee Guide)}.
\newblock Berklee Press, 2000.

\bibitem[Piramuthu(2004)]{piramuthu2004evaluating}
S.~Piramuthu.
\newblock Evaluating feature selection methods for learning in data mining
  applications.
\newblock \emph{European journal of operational research}, 156\penalty0
  (2):\penalty0 483--494, 2004.

\bibitem[Quinlan(1993)]{quinlan1993}
J.R. Quinlan.
\newblock \emph{C4. 5: programs for machine learning}, volume~1.
\newblock Morgan kaufmann, 1993.

\bibitem[Rish(2001)]{rish2001empirical}
I.~Rish.
\newblock An empirical study of the naive bayes classifier.
\newblock In \emph{IJCAI 2001 workshop on empirical methods in artificial
  intelligence}, volume~3, pages 41--46, 2001.

\bibitem[Ruggieri(2002)]{ruggieri2002efficient}
S.~Ruggieri.
\newblock Efficient c4. 5 [classification algorithm].
\newblock \emph{Knowledge and Data Engineering, IEEE Transactions on},
  14\penalty0 (2):\penalty0 438--444, 2002.

\bibitem[Salganik et~al.(2006)Salganik, Dodds, and
  Watts]{salganik2006experimental}
M.J. Salganik, P.S. Dodds, and D.J. Watts.
\newblock Experimental study of inequality and unpredictability in an
  artificial cultural market.
\newblock \emph{science}, 311\penalty0 (5762):\penalty0 854--856, 2006.

\bibitem[Schindler and Rauber(2012)]{schindler2012capturing}
A.~Schindler and A.~Rauber.
\newblock Capturing the temporal domain in echonest features for improved
  classification effectiveness.
\newblock \emph{Proc. Adaptive Multimedia Retrieval.(Oct. 2012)}, 2012.

\bibitem[Schnitzer et~al.(2009)Schnitzer, Flexer, and
  Widmer]{schnitzer2009filter}
D.~Schnitzer, A.~Flexer, and G.~Widmer.
\newblock A filter-and-refine indexing method for fast similarity search in
  millions of music tracks.
\newblock In \emph{Proceedings of the 10th International Conference on Music
  Information Retrieval (ISMIR09)}, 2009.

\bibitem[Shneiderman(2002)]{shneiderman2002inventing}
B.~Shneiderman.
\newblock Inventing discovery tools: combining information visualization with
  data mining1.
\newblock \emph{Information Visualization}, 1\penalty0 (1):\penalty0 5--12,
  2002.

\bibitem[Tan et~al.(2007)]{tan2007introduction}
P.N. Tan et~al.
\newblock \emph{Introduction to data mining}.
\newblock Pearson Education India, 2007.

\bibitem[Todd(1992)]{todd1992dynamics}
N.P.M. Todd.
\newblock The dynamics of dynamics: A model of musical expression.
\newblock \emph{The Journal of the Acoustical Society of America}, 91:\penalty0
  3540, 1992.

\bibitem[Tu(1996)]{tu1996advantages}
J.V. Tu.
\newblock Advantages and disadvantages of using artificial neural networks
  versus logistic regression for predicting medical outcomes.
\newblock \emph{Journal of clinical epidemiology}, 49\penalty0 (11):\penalty0
  1225--1231, 1996.

\bibitem[Tzanetakis and Cook(2002)]{tzanetakis2002musical}
G.~Tzanetakis and P.~Cook.
\newblock Musical genre classification of audio signals.
\newblock \emph{Speech and Audio Processing, IEEE transactions on}, 10\penalty0
  (5):\penalty0 293--302, 2002.

\bibitem[Vapnik(1995)]{vapnik}
V.N. Vapnik.
\newblock \emph{The nature of statistical learning theory}.
\newblock Springer-Verlag New York, Inc., New York, NY, USA, 1995.

\bibitem[Webb(1999)]{Webb1999}
J.~Webb.
\newblock \emph{Tunesmith: Inside the Art of Songwriting}.
\newblock Hyperion, 1999.

\bibitem[Weka(2013{\natexlab{a}})]{grid2013}
Weka.
\newblock Weka documentation, class gridsearch, 2013{\natexlab{a}}.
\newblock URL
  \url{http://weka.sourceforge.net/doc.stable/weka/classifiers/meta/GridSearch.html}.

\bibitem[Weka(2013{\natexlab{b}})]{wikiweka2013}
Weka.
\newblock Optimizing parameters, 2013{\natexlab{b}}.
\newblock URL \url{http://weka.wikispaces.com/Optimizing+parameters}.

\bibitem[Whitman and Smaragdis(2002)]{whitman2002combining}
B.~Whitman and P.~Smaragdis.
\newblock Combining musical and cultural features for intelligent style
  detection.
\newblock In \emph{Proc. Int. Symposium on Music Inform. Retriev.(ISMIR)},
  pages 47--52, 2002.

\bibitem[Wiginton(1980)]{wiginton1980note}
J.C. Wiginton.
\newblock A note on the comparison of logit and discriminant models of consumer
  credit behavior.
\newblock \emph{Journal of Financial and Quantitative Analysis}, 15\penalty0
  (03):\penalty0 757--770, 1980.

\bibitem[Witten and Frank(2005)]{witten2005data}
I.H. Witten and E.~Frank.
\newblock \emph{Data Mining: Practical machine learning tools and techniques}.
\newblock Morgan Kaufmann, 2005.

\bibitem[Wolberg and Mangasarian(1990)]{wolberg1990multisurface}
W.H. Wolberg and O.L. Mangasarian.
\newblock Multisurface method of pattern separation for medical diagnosis
  applied to breast cytology.
\newblock \emph{Proceedings of the national academy of sciences}, 87\penalty0
  (23):\penalty0 9193--9196, 1990.

\bibitem[Wu et~al.(2008)Wu, Kumar, Ross~Quinlan, Ghosh, Yang, Motoda,
  McLachlan, Ng, Liu, Yu, et~al.]{wu2008top}
X.~Wu, V.~Kumar, J.~Ross~Quinlan, J.~Ghosh, Q.~Yang, H.~Motoda, G.J. McLachlan,
  A.~Ng, B.~Liu, P.S. Yu, et~al.
\newblock Top 10 algorithms in data mining.
\newblock \emph{Knowledge and Information Systems}, 14\penalty0 (1):\penalty0
  1--37, 2008.

\end{thebibliography}

% 
%  \begin{frame}[plain]
%  \frametitle{T3mean (x) versus beatdif var (y)}
% \includegraphics[scale=0.65]{Xt3meanYbeatdifvar.jpg}
%  \end{frame}
%  
%   \begin{frame}[plain]
%    \frametitle{T1 median (x) vs T3 mean (y)}
% \includegraphics[scale=0.65]{T1xmedianT3ymean.jpg}
%  \end{frame}
%  

\end{document}